\documentclass[useAMS,usenatbib]{mn2e}
\usepackage{graphicx,amsmath,amssymb}

\newcommand{\kmps}{\ensuremath{\mathrm{km\ s}^{-1}}}

\newcommand{\Reff}{\ensuremath{R_{\mathrm{eff}}}}

\newcommand{\figeximages}{
  \begin{figure}
    \includegraphics[width=3in]{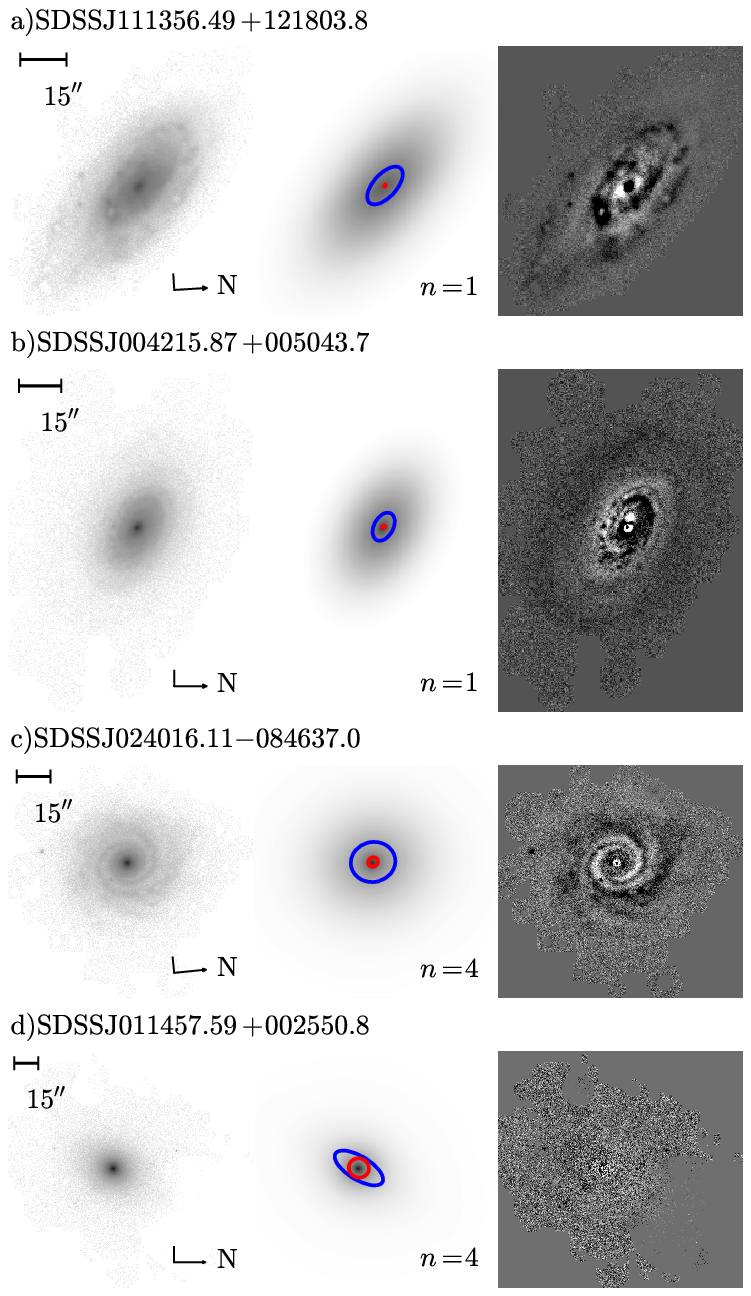}
    \caption{Images of fitted galaxies. The first column is the SDSS
      $r-$band image, the center is the model fit with an $n=(1,4)$ bulge, and
      the final column is the residual. The blue ellipse contains 1/2 the
      light from the disk, while the red ellipse contains half the light
      from the bulge.}
    \label{fig:example_fits}
  \end{figure}
}
\newcommand{\figexprof}{
  \begin{figure}
    \includegraphics[width=3in]{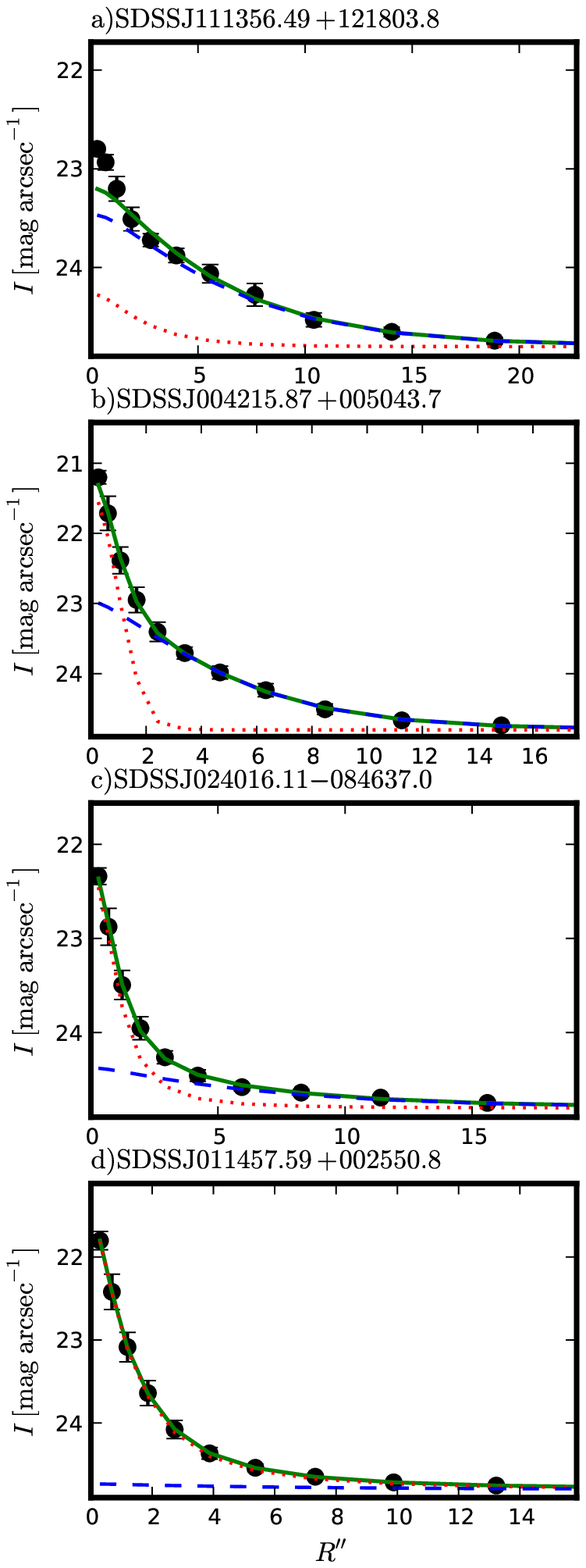}
    \caption{The one-dimensional profiles of the galaxies in
      Figure \ref{fig:example_fits}. The axis along which the profile is
      taken is the semimajor axis of the bulge or disk, depending on which
      has a larger \Reff. The surface brightness is the mean surface
      brightness in elliptical annuli with
      axis ratios and position angles taken from the bulge or disk. The
      points are the surface brightness of the image with  errorbars denoting the
      standard deviation of the flux in an annulus. The green solid line
      is the model flux. The blue dashed line is the disk flux and the
      red dotted line is the bulge flux.}
    \label{fig:example_profs}
  \end{figure}
}

\newcommand{\figcolourBTT}{
  \begin{figure*}
    \includegraphics[width=6.7in]{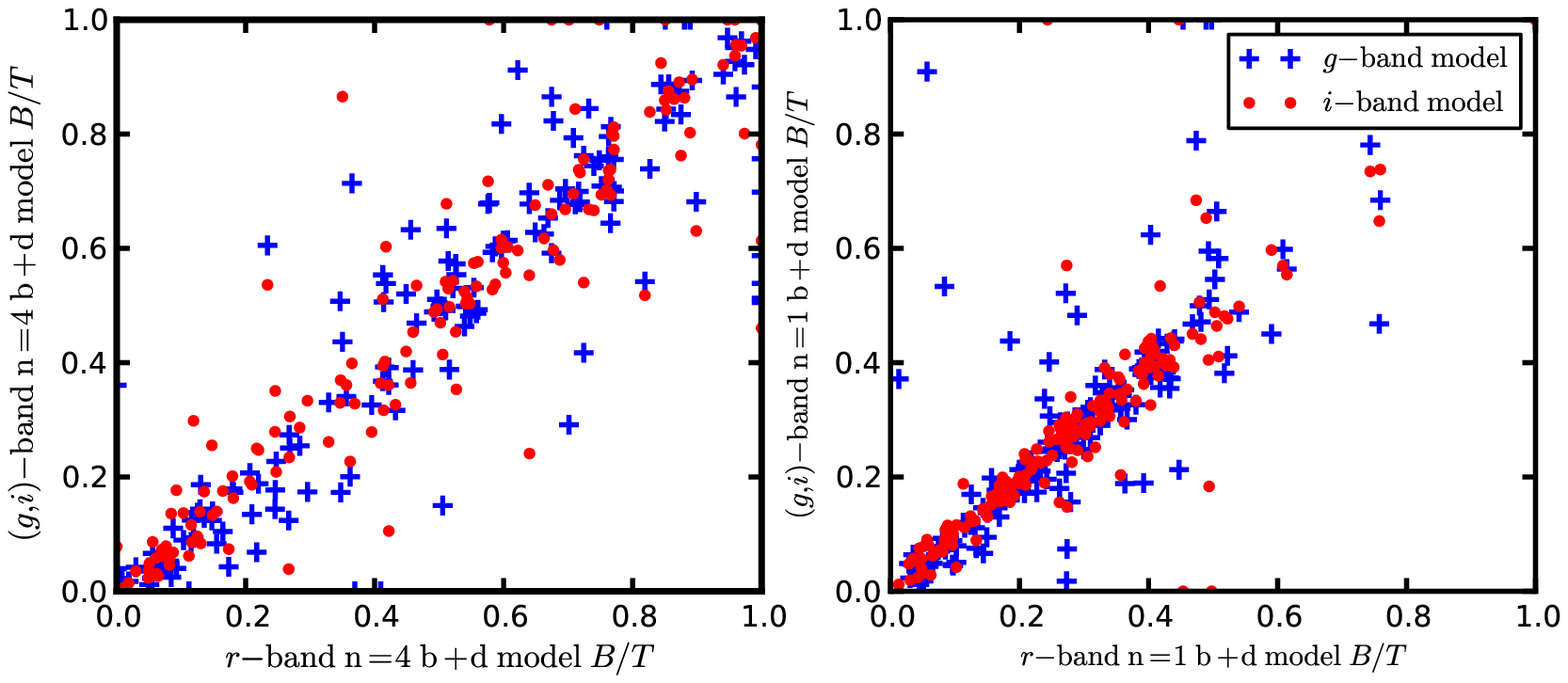}
    \caption{Comparison of the $B/T$ for an $n_b=4$ B+D model (left)
      and an $n_b=1$ B+D model (right) in the $r-$band
      for galaxies fit in the $g$,$r$,and $i$ bands. The
      $x-$axis is the $B/T$ of the model fit in $r$. The blue crosses
      show the $B/T$ in $r$ measured by scaling the model fit to the
      $g$-band. The red points are the same for the $i$-band.}  
    \label{fig:btt_colour_comp}
  \end{figure*}
}

\newcommand{\figcolourReff}{
  \begin{figure}
    \includegraphics[width=3in]{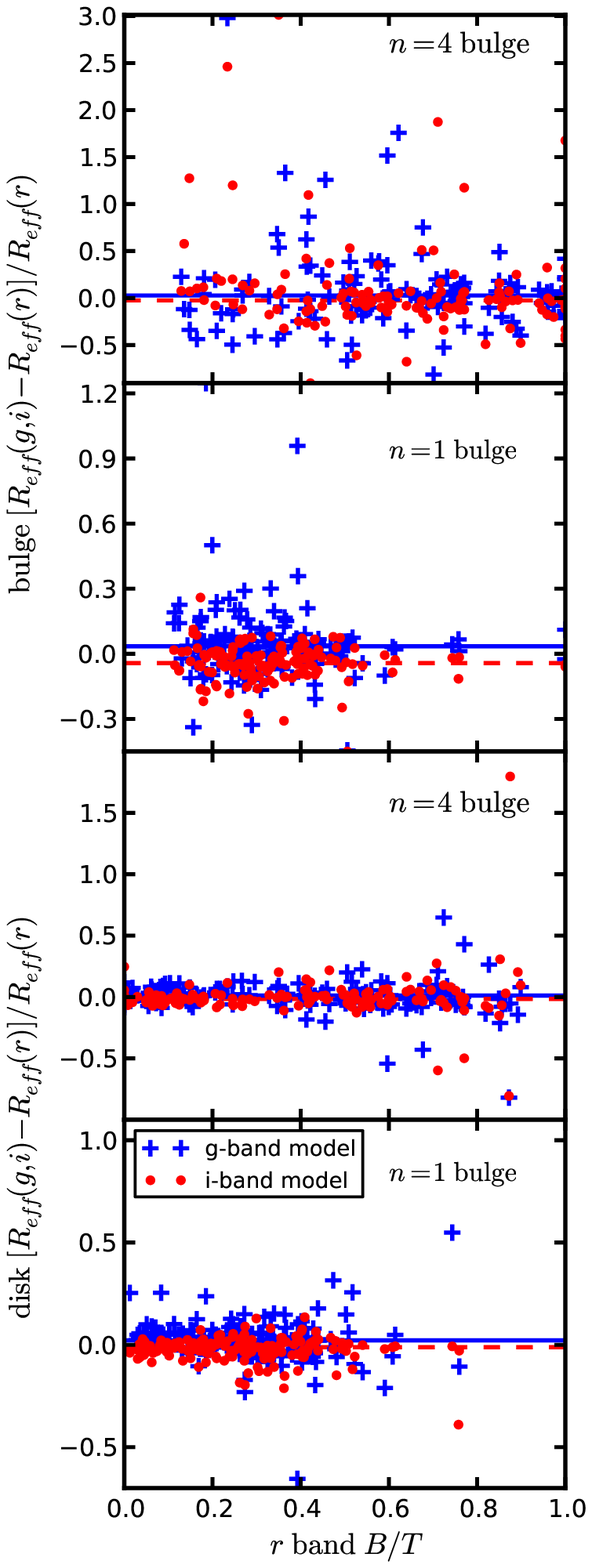}
    \caption{Comparison of the scale lengths fit in $g$ and $i$ to
      those found using the $r$-band images. The top two figures show
      the relative difference in bulge \Reff{} for an $n_b=4$ B+D
      (top) and an $n_b=1$ B+D (second). Galaxies with $B/T < 0.1$ in
      any band are not shown. The 
      bottom two figures show the same for the the disk \Reff, with
      galaxies with $B/T > 0.9$ in any band removed. The blue solid
      (red dashed) lines show the median difference between the $g$
      ($i$) and $r$ band fits.}
    \label{fig:reff_colour_comp}
  \end{figure}
}

\newcommand{\figsnresNReff}{
 \begin{figure}
    \includegraphics[width=3.2in]{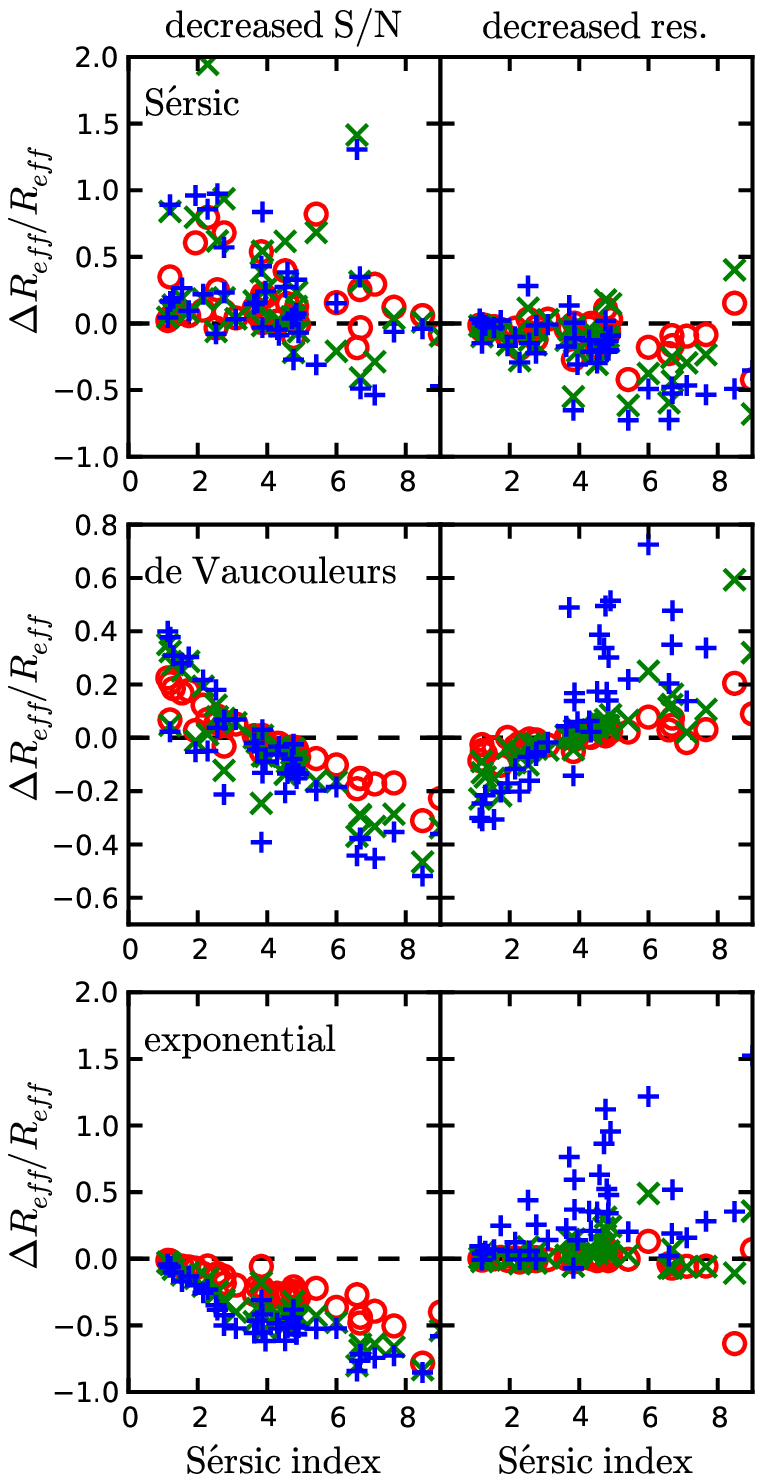}
    \caption{Fractional change in measured \Reff{} as a function of
      S\'ersic index for Sample A galaxies.  The red circles show an
      increase in 
      the sky noise by a factor of 2 (left), and increase the PSF width by a
      factor of 2 (right). The green $\times$s and the blue crosses
      show increases in 
      the sky noise and the PSF width by a factor of 4 and 10,
      respectively. These changes are equivalent to the changes that
      would occur in resolution and S/N if the galaxy were moved
      farther away by a factors of 2, 4, and 10 (in a flat cosmology). }
    \label{fig:NReff_SNRES_comp}
  \end{figure}
}
\newcommand{\figsnresSersicN}{
 \begin{figure*}
    \includegraphics[width=6in]{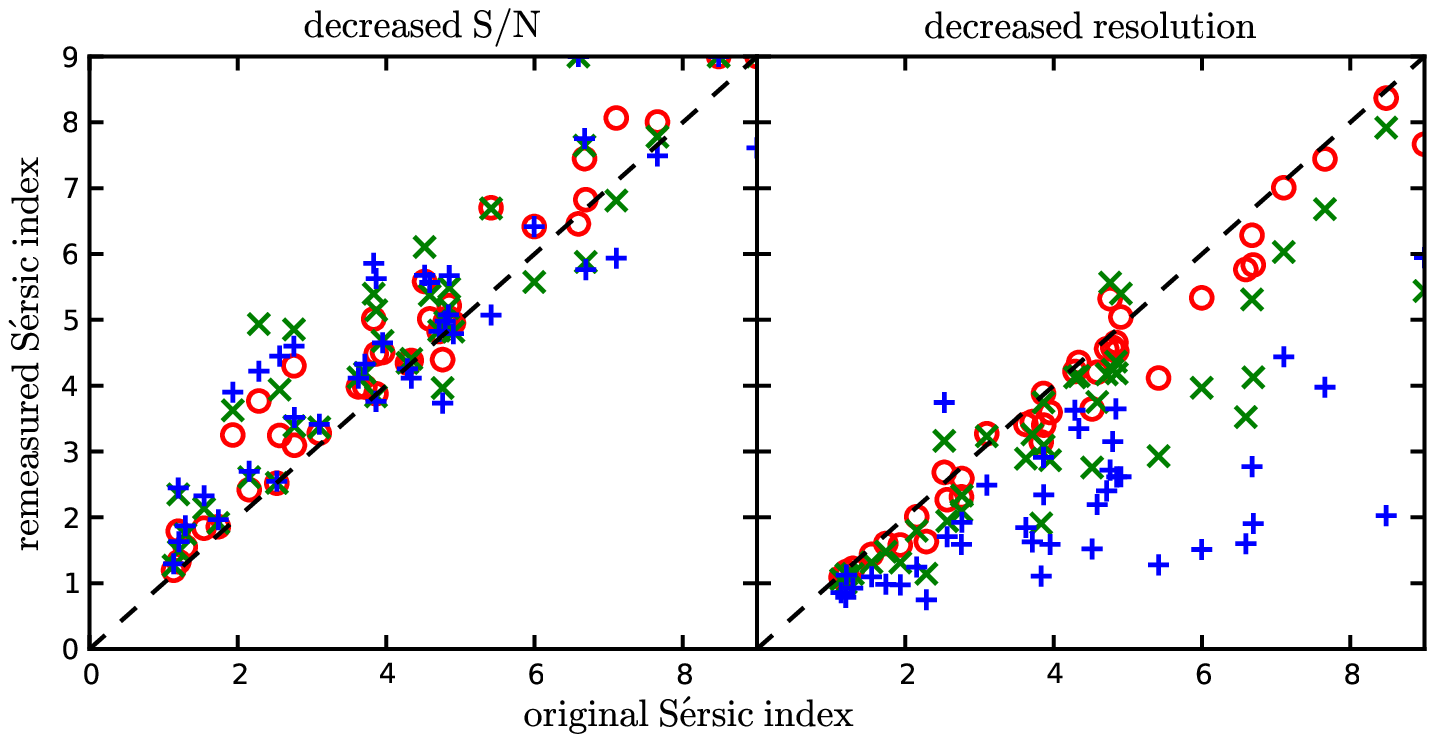}
    \caption{Comparison of the S\'ersic index fit to Sample A galaxies
      while decreasing S/N (left) and resolution (right). The symbols
      are the same as in Figure \ref{fig:NReff_SNRES_comp}. }
    \label{fig:N_SNRES_comp}
  \end{figure*}
}
\newcommand{\figsnresNMag}{
 \begin{figure}
    \includegraphics[width=3.2in]{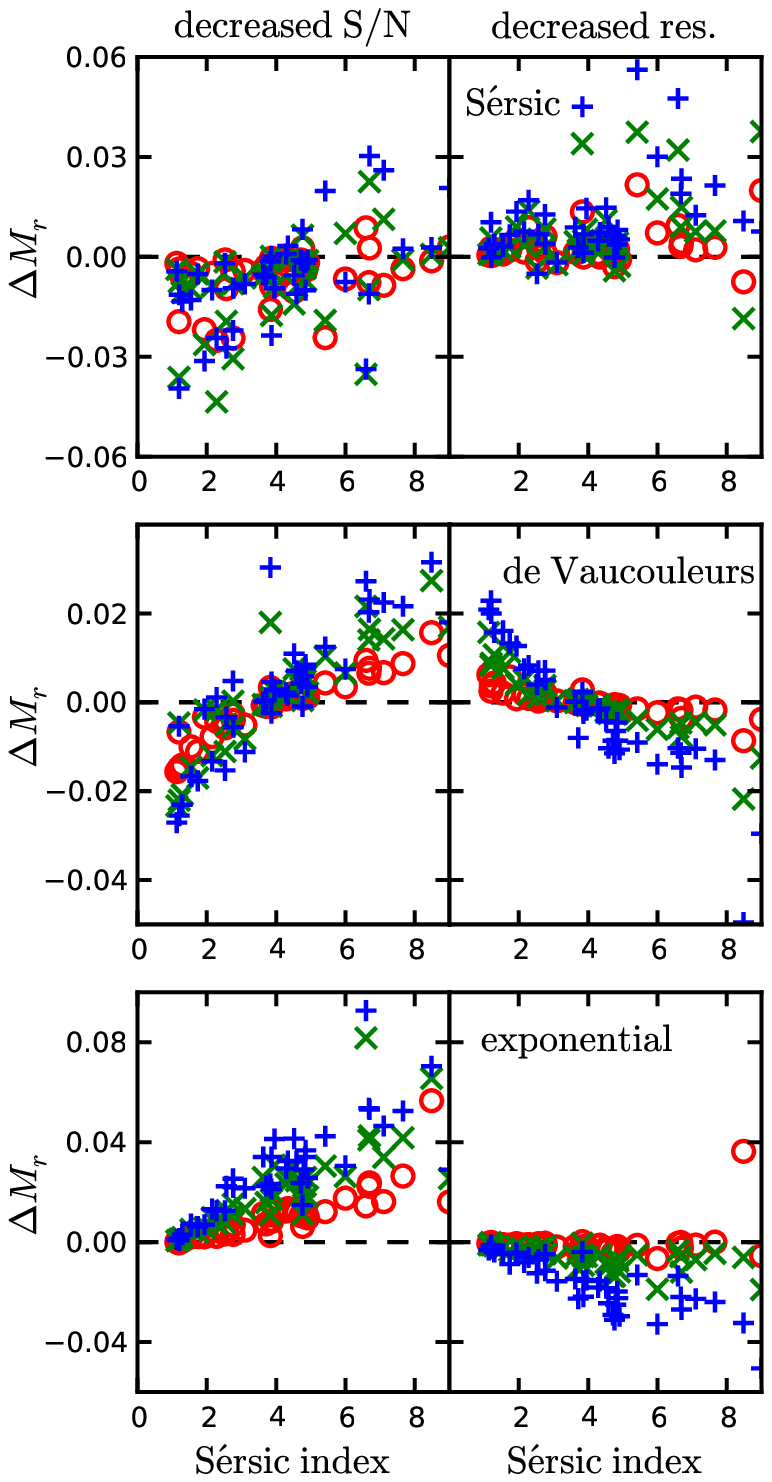}
    \caption{Change in measured $r$-band magnitude with
      decreased S/N(left) and resolution(right) as a
      function of S\'ersic index for Sample A galaxy images. The symbols
      are as in Figure \ref{fig:NReff_SNRES_comp}. }
    \label{fig:NMag_SNRES_comp}
  \end{figure}
}
\newcommand{\figsnresBTTMag}{
 \begin{figure}
    \includegraphics[width=3.2in]{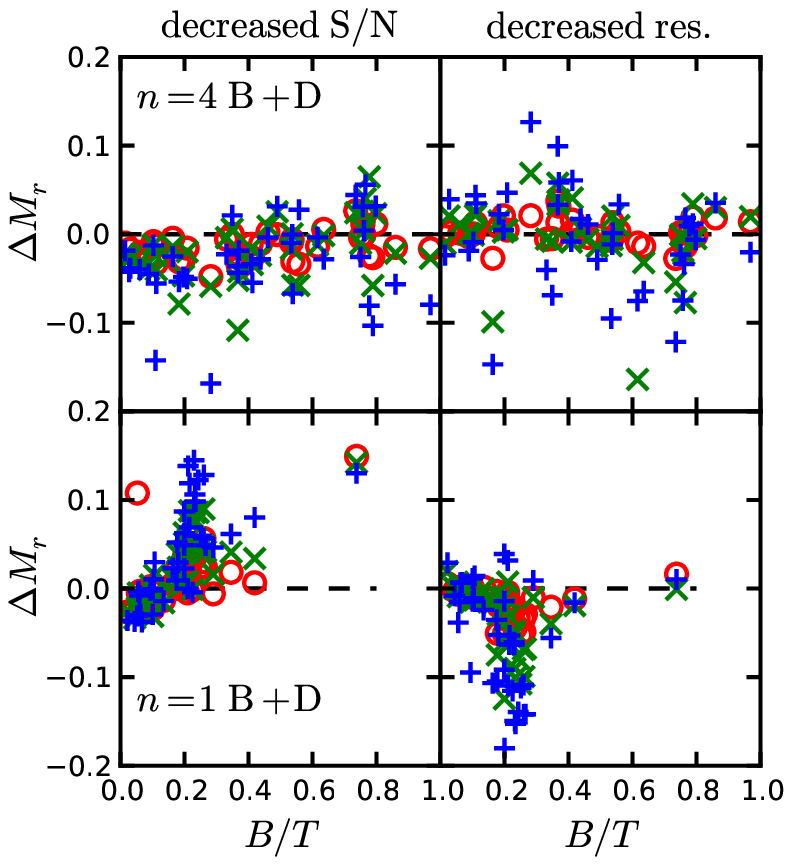}
    \caption{Change in measured $r$-band magnitude for B+D models with
      decreased S/N(left) and resolution(right) as a
      function of $B/T$ for Sample A galaxy images. The symbols are as
      in Figure \ref{fig:NReff_SNRES_comp}.}
    \label{fig:BTTMag_SNRES_comp}
  \end{figure}
}

\newcommand{\figsnresBTT}{
 \begin{figure}
    \includegraphics[width=3.1in]{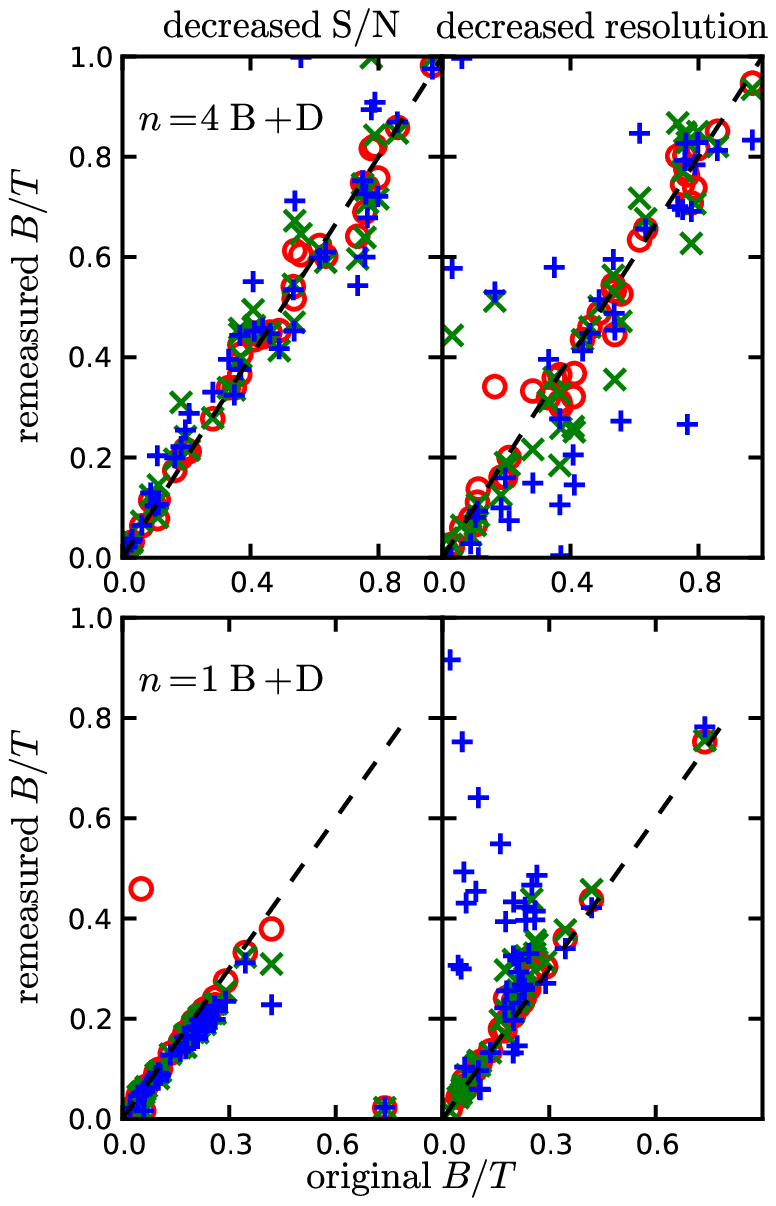}
    \caption{Change in $B/T$ with decreased S/N(left) and resolution
      (right) as a function of $B/T$ measured in the non-degraded
      images for Sample A galaxies. The symbols are as in Figure
      \ref{fig:NReff_SNRES_comp}.} 
    \label{fig:BTT_SNRES_comp}
  \end{figure}
}

\newcommand{\figSNRESRbulge}{
 \begin{figure}
    \includegraphics[width=3.1in]{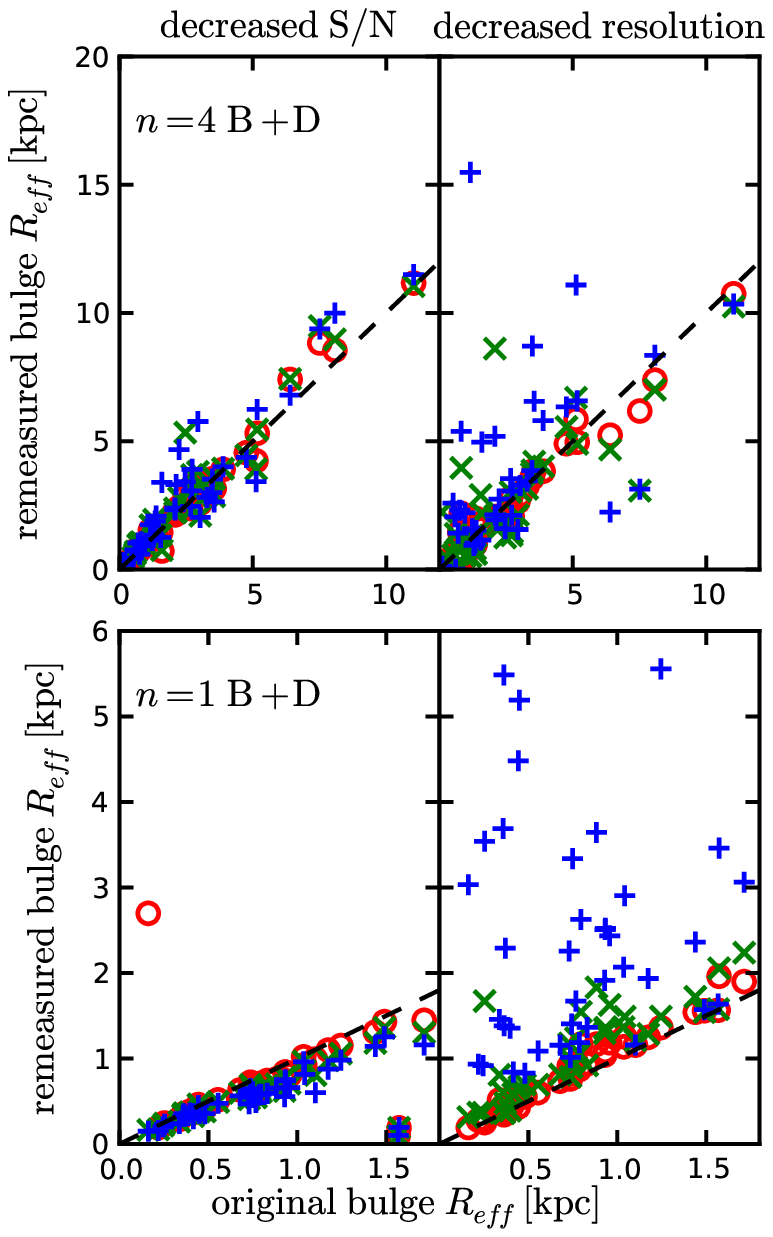}
    \caption{Change in measured bulge \Reff{} for the $n_b=4$ B+D models
      (top) and $n_b=1$ B+D models (bottom) as a function of bulge
      \Reff{}. The left(right) column 
      shows trends with decreased S/N(resoltuion). The symbols
      are as in Figure \ref{fig:NReff_SNRES_comp}. }
    \label{fig:Rebulge_SNRES_comp}
  \end{figure}
}

\newcommand{\figSNRESRdisk}{
 \begin{figure}
   \includegraphics[width=3.2in]{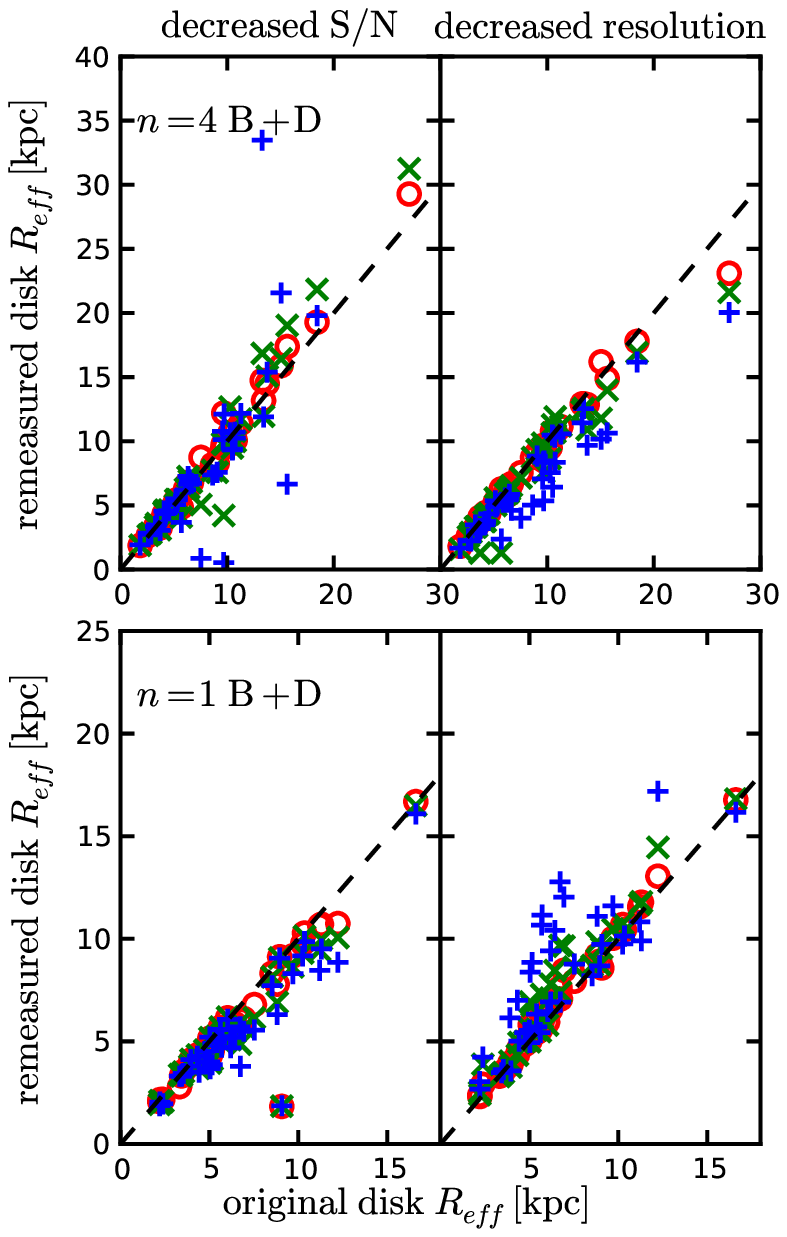}
   \caption{Same as Figure \ref{fig:Rebulge_SNRES_comp} for the disk
     \Reff{}s.}
    \label{fig:Redisk_SNRES_comp}
  \end{figure}
}

\newcommand{\figSNRESexA}{
 \begin{figure*}
    \includegraphics[width=6.7in]{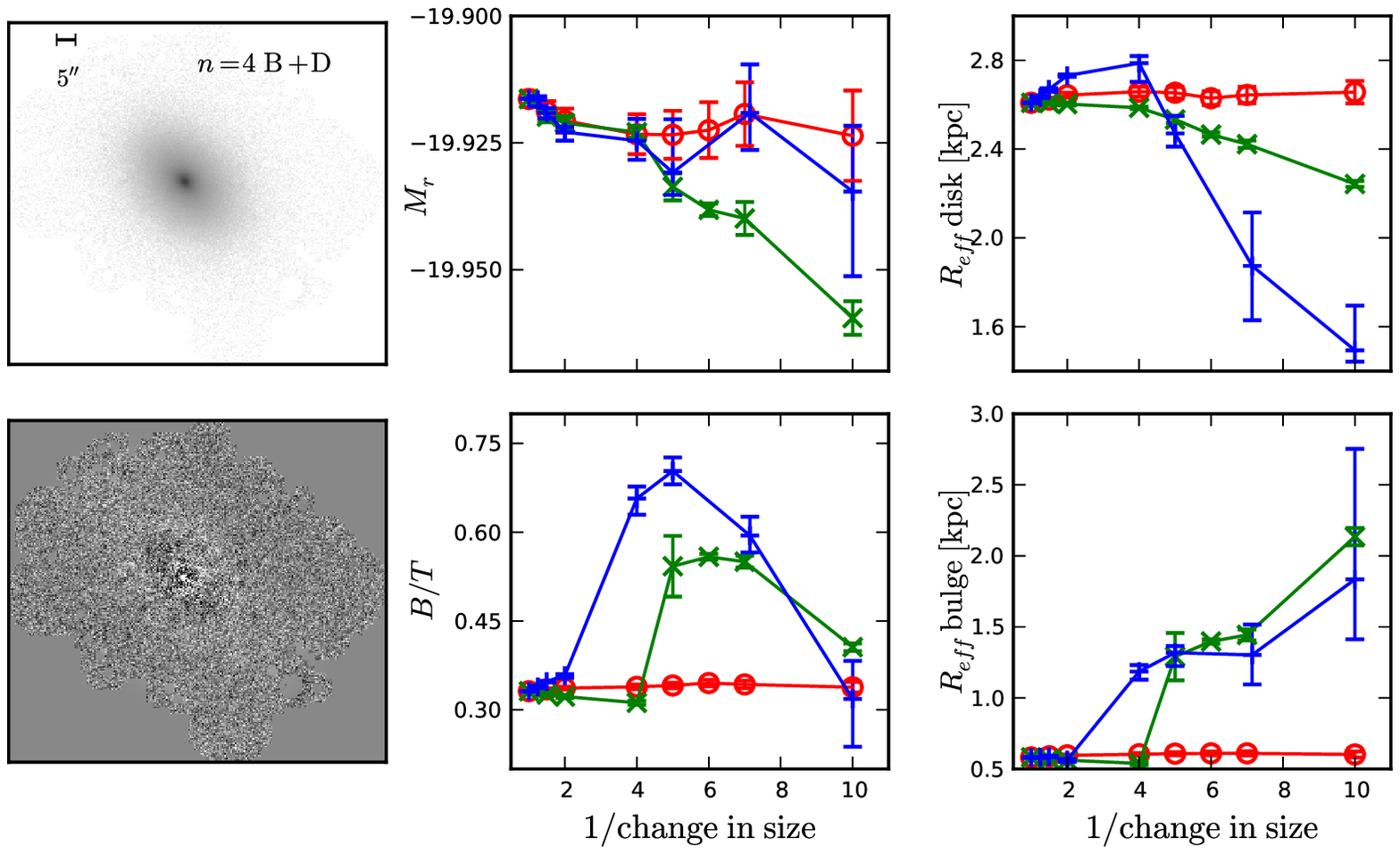}
    \caption{Change in model parameters for the galaxy pictured. The
      model used is named in the first panel. The lower left panel
      shows the residuals (black is high and positive). The blue lines
      with crosses show the changes with increasing redshift,
      i.e. $x=2$ corresponds to an increase in redshift such that the
      angular size decreases by a factor of $2$. The green $\times$s
      show the changes with 
      decreasing resolution, i.e. at $x=2$ the image is convolved with
      a PSF twice as large as the original. The red circles show the
      changes with decreasing S/N, i.e. at $x=2$, the noise is
      increased by a factor of $2$ as it would be if the galaxy were
      moved twice as far away. }
    \label{fig:SNRES_exA}
  \end{figure*}
}
\newcommand{\figSNRESexB}{
 \begin{figure*}
    \includegraphics[width=6.7in]{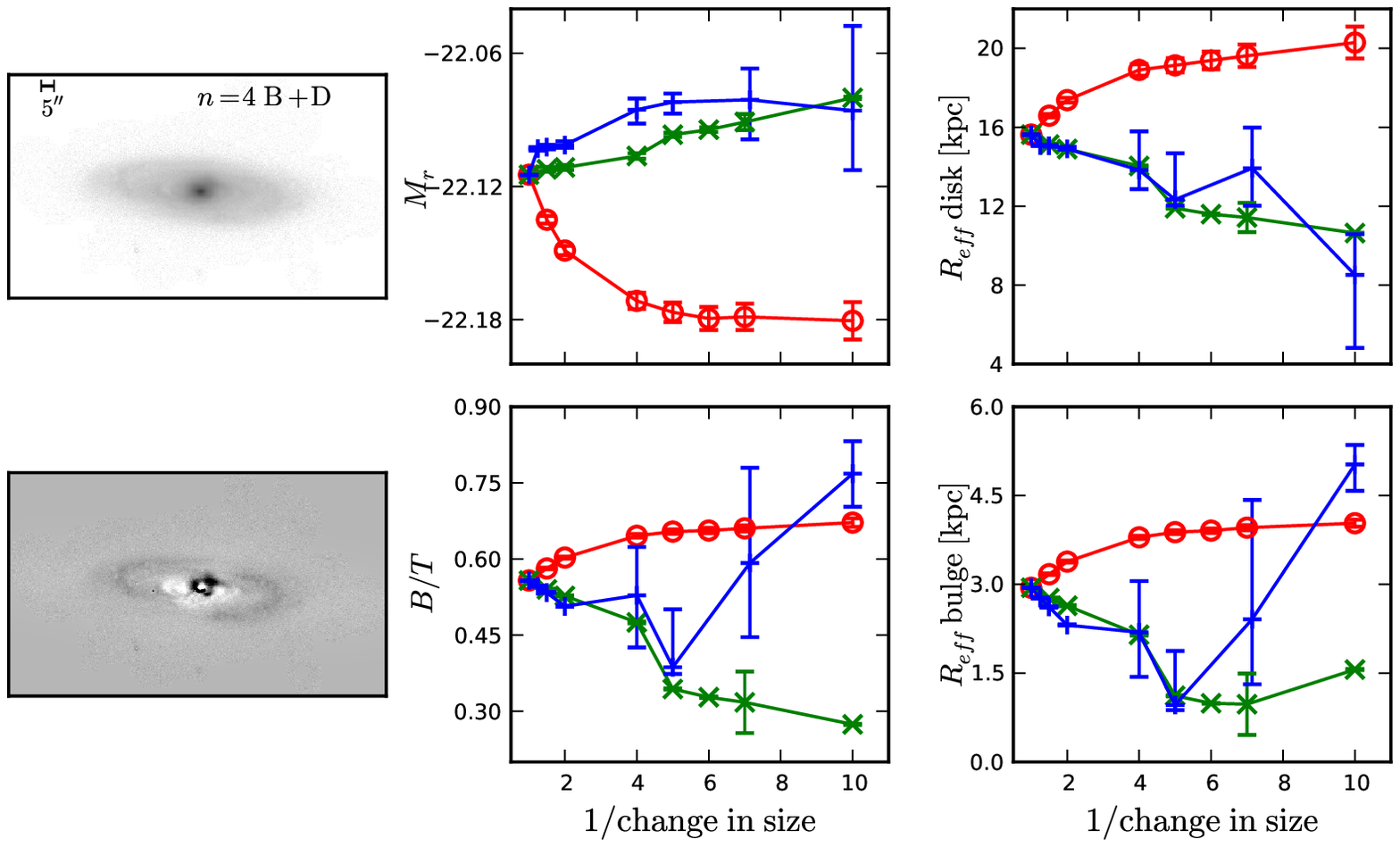}
    \caption{See Figure \ref{fig:SNRES_exA}.}
    \label{fig:SNRES_exB}
  \end{figure*}
}
\newcommand{\figSNRESexC}{
 \begin{figure*}
    \includegraphics[width=6.7in]{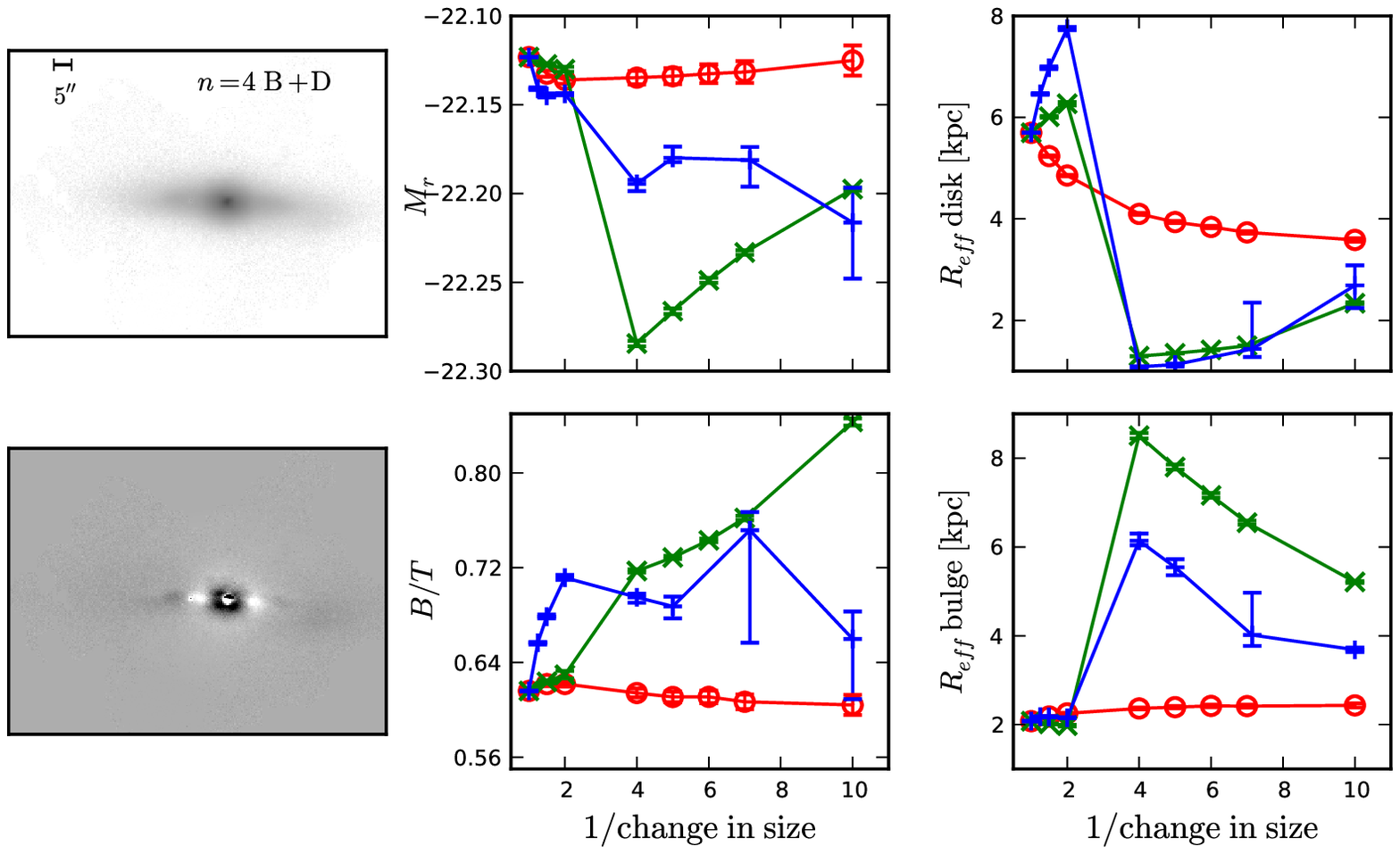}
    \caption{See Figure \ref{fig:SNRES_exA}.}
    \label{fig:SNRES_exC}
  \end{figure*}
}
\newcommand{\figSNRESexD}{
 \begin{figure*}
    \includegraphics[width=6.7in]{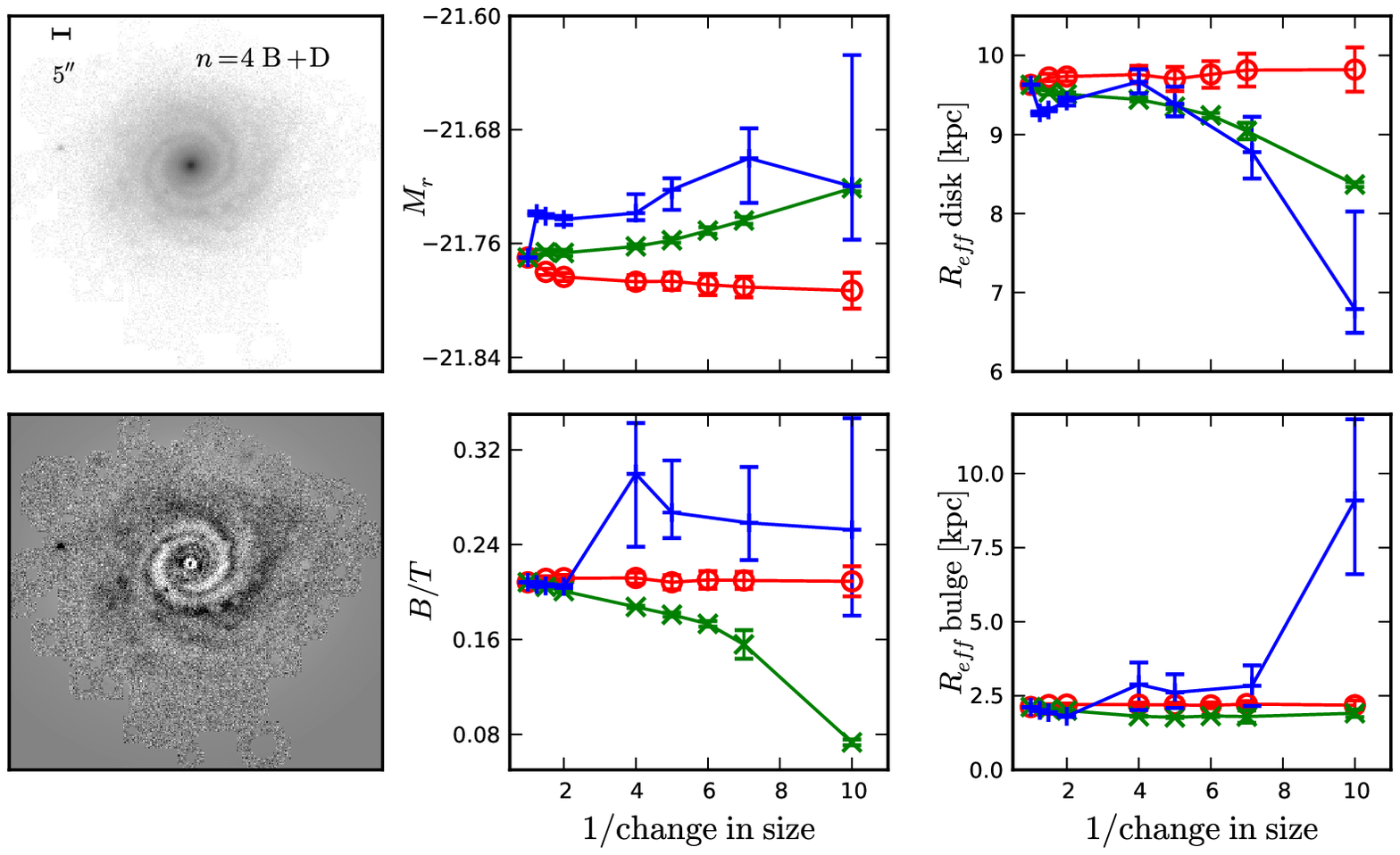}
    \caption{See Figure \ref{fig:SNRES_exA}.}
    \label{fig:SNRES_exD}
  \end{figure*}
}

\newcommand{\figZNMag}{
 \begin{figure}
    \includegraphics[width=3.in]{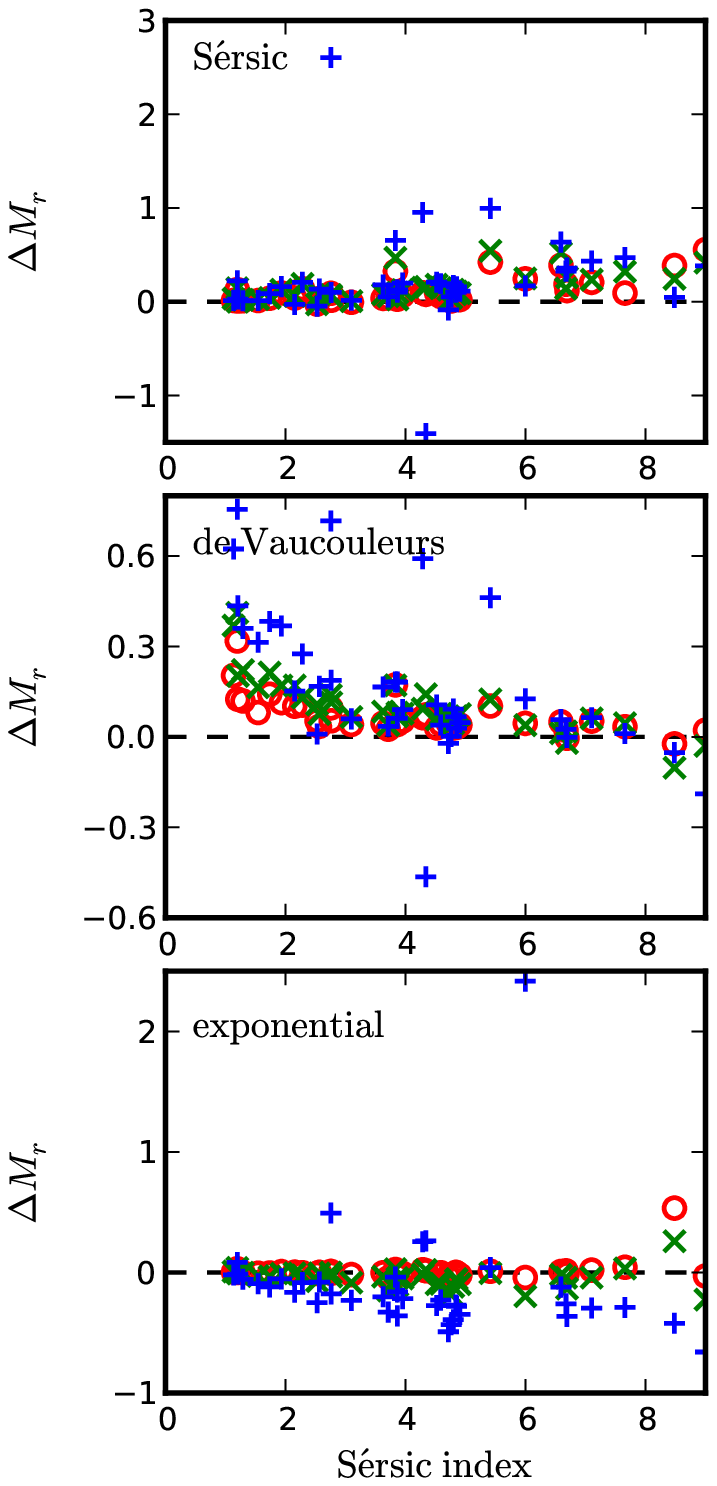}
    \caption{Change in $r$-band magnitude with increasing redshift as
      a function of S\'ersic index for Sample A galaxies. The symbols 
      are as in Figure \ref{fig:NReff_SNRES_comp}, with the red circles,
      green $\times$s and blue crosses representing a decrease in
      angular size by a factor of $2$, $4$, and $10$,
      respectively. These plots can be compared directly to the results in
      Figure \ref{fig:NMag_SNRES_comp}.}
    \label{fig:NMag_Z_comp}
  \end{figure}
}

\newcommand{\figZSersicN}{
 \begin{figure}
    \includegraphics[width=3in]{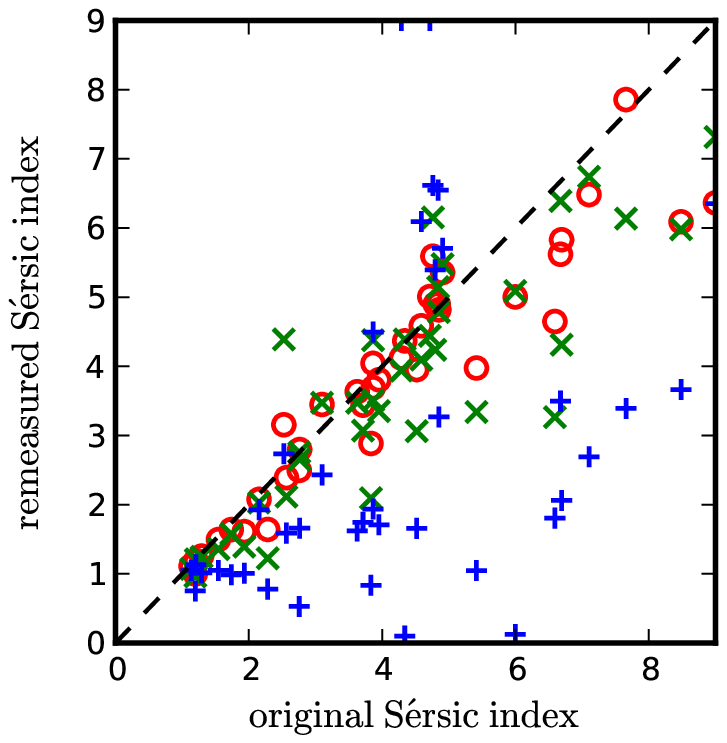}
    \caption{Comparison of the S\'ersic index measured for redshifted Sample
      A galaxy images to the S\'ersic index measured for the original
      Sample A images. The symbols are as in Figure
      \ref{fig:NMag_SNRES_comp}. }
    \label{fig:N_Z_comp}
  \end{figure}
}

\newcommand{\figZBTT}{
 \begin{figure}
    \includegraphics[width=3.in]{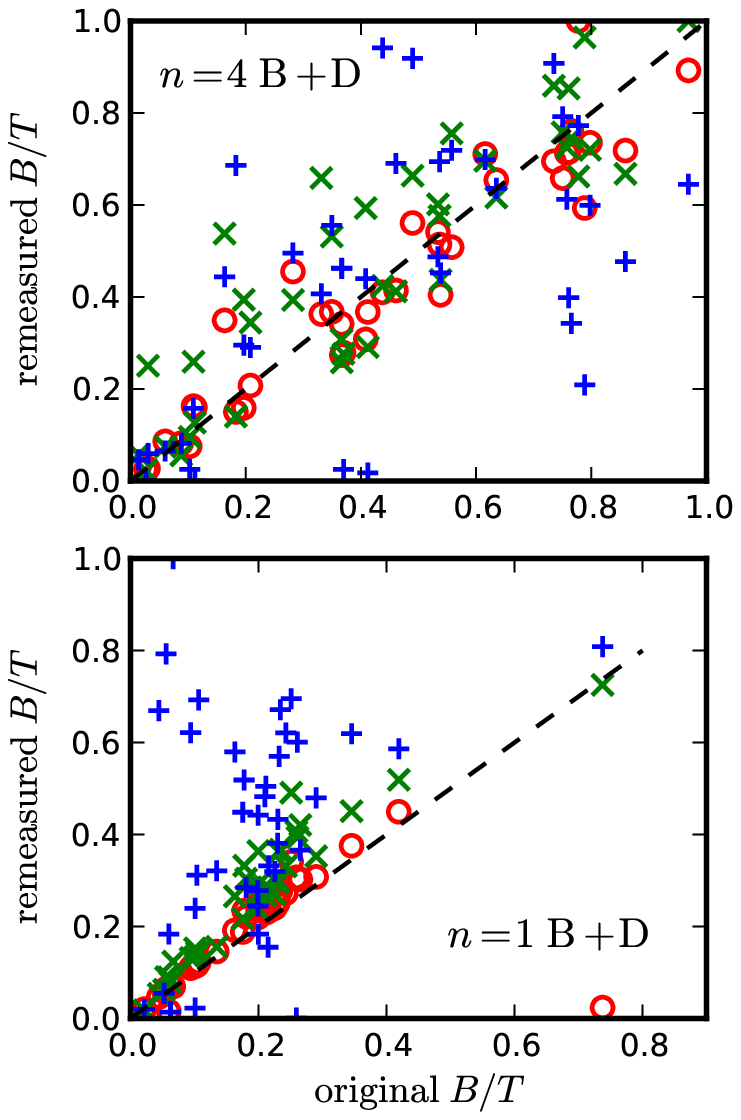}
    \caption{Comparison of the B/T measured for redshifted Sample
      A galaxy images to the B/T measured for the original
      Sample A images. The symbols are as in Figure
      \ref{fig:NMag_SNRES_comp}. }
    \label{fig:BTT_Z_comp}
  \end{figure}
}

\newcommand{\figdVcQ}{
 \begin{figure}
    \includegraphics[width=3.in]{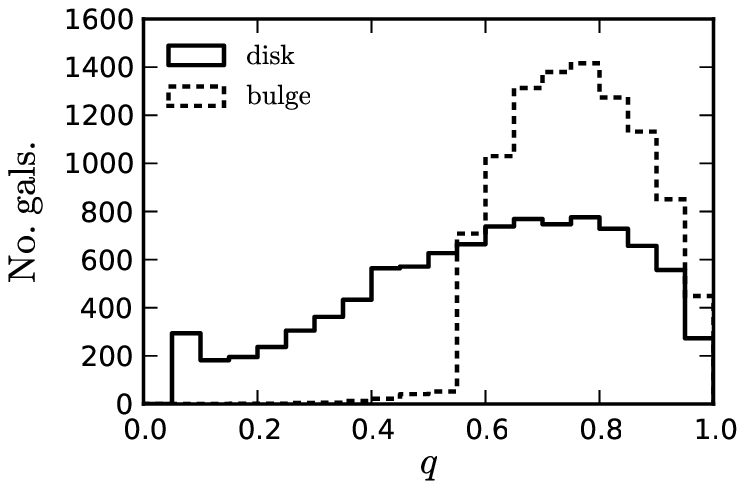}
    \caption{Distribution of $n_b=4$ B+D bulge and disk
      axis ratios for the $9,692$ galaxies best fit with a single
      component de Vaucouleurs model.}
    \label{fig:dvc_q}
  \end{figure}
}

\newcommand{\figBulgeGR}{
 \begin{figure}
    \includegraphics[width=3.in]{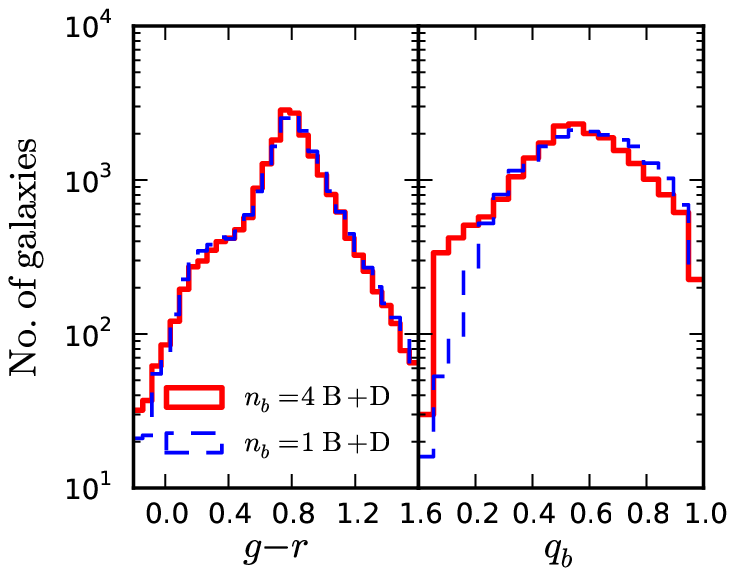}
    \caption{Distribution of bulge colours and axis ratios for all
      $20,726$ B+D galaxies, fit with both B+D models.}
    \label{fig:bulgeGR}
  \end{figure}
}
\newcommand{\figCatN}{
 \begin{figure}
    \includegraphics[width=3.in]{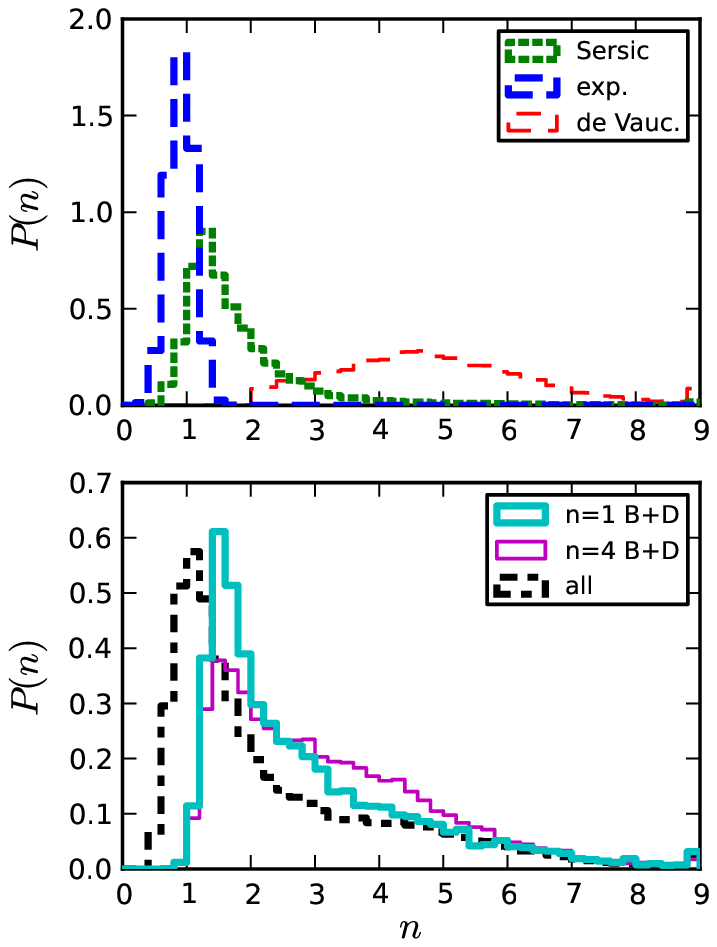}
    \caption{Distribution of S\'ersic indices (from the single
      component S\'ersic fits) for the galaxies in different categories. The
      galaxies are placed in the five categories according to the
      alogrithm described in \S\ref{ssec:n1n4bulge}. Note the
      different scales in the two plots. }
    \label{fig:Cat_N}
  \end{figure}
}

\newcommand{\figCatMag}{
 \begin{figure*}
    \includegraphics[width=6.7in]{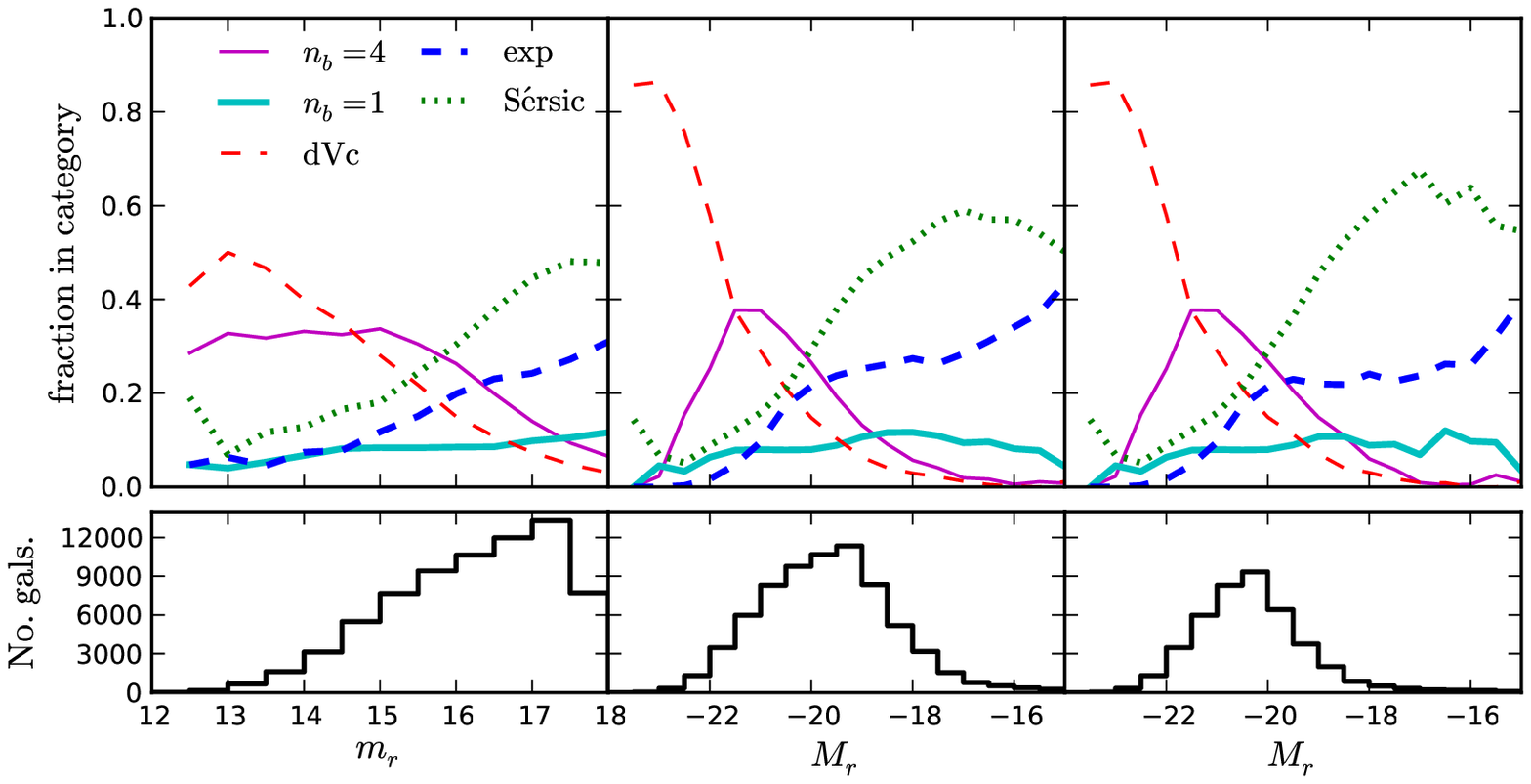}
    \caption{Top panels: Fraction of galaxies in each category as a function of
      apparent magnitude (left) and absolute magnitude (middle and
      right). Lower panels: Apparent and absolute magnitude
      distributions for the sample in each of the top panels. The rightmost
      column shows galaxies with $m_r < 16.7$. }
    \label{fig:Cat_Mag}
  \end{figure*}
}

\newcommand{\figKormendy}{
 \begin{figure}
    \includegraphics[width=3.in]{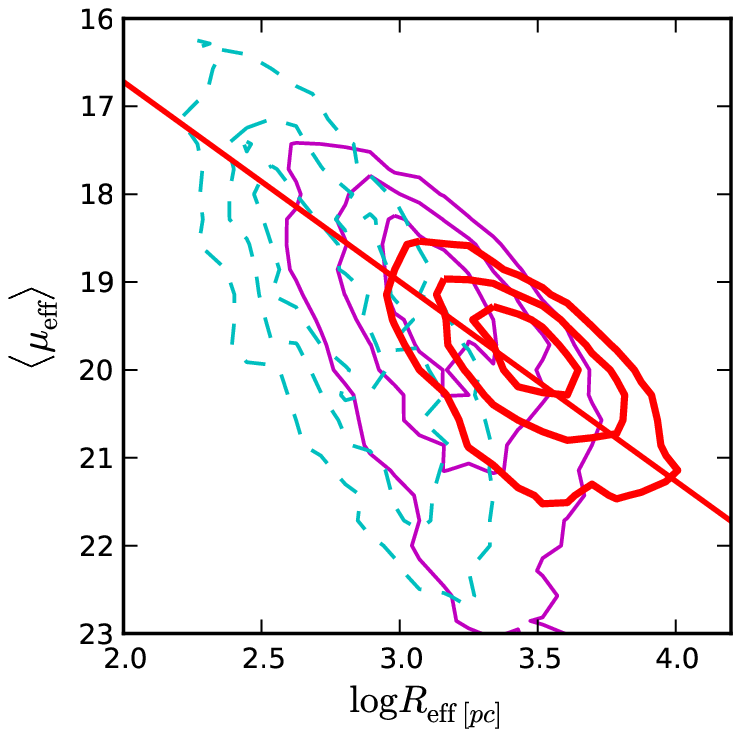}
    \caption{Kormendy relation for bulges of $n_b=4$ B+D (thin solid, magenta),
      $n_b=1$ B+D (dashed, cyan) and de Vaucouleurs(thick solid, red)
      modeled galaxies. The straigt line is a fit to the binned
      medians of the elliptical galaxies, and is given by the formula $\langle
      \mu_{\mathrm{eff}}\rangle = (2.27\pm0.04)\log\Reff\mathrm{[pc]}
      + (12.18\pm 0.15)$. The contours enclose $30\%$, $50\%$, and
      $80\%$ of each category. } 
    \label{fig:kormendy}
  \end{figure}
}

\newcommand{\figdiskIncDist}{
 \begin{figure}
   \includegraphics[width=3.in]{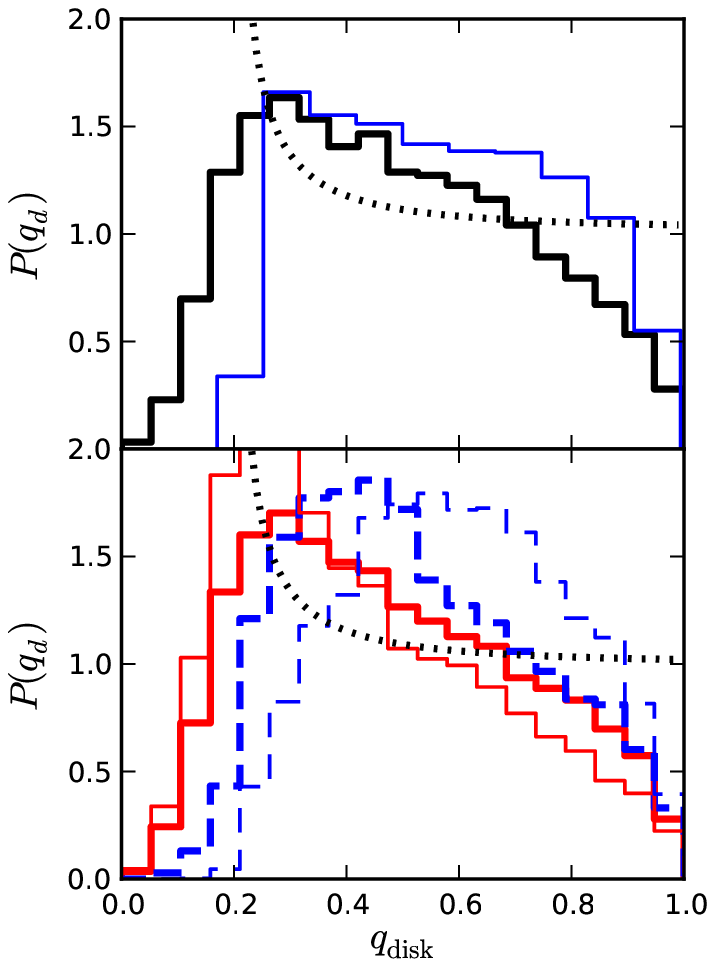}
    \caption{Distribution of disk axis ratios. The dotted line in both
    panels show the expected distribution for a constant disk
    flattening of $5$. Top panel: The thick black line is the
    distribution for all disks (from exponential and B+D galaxies). The thin 
    blue line is for single component exponential fits. Bottom panel:
    The thick red, solid (blue, dashed) line is for all the B+D
    galaxies fit with an $n_b=4$ ($n_b=1$) B+D model. The thin lines of
    the same colour/style only show the galaxies which were
    categorized as $n_b=4$ B+D or $n_b=1$ B+D.}
    \label{fig:diskIncDist}
  \end{figure}
}

\newcommand{\figMallerExp}{
 \begin{figure*}
    \includegraphics[width=6in]{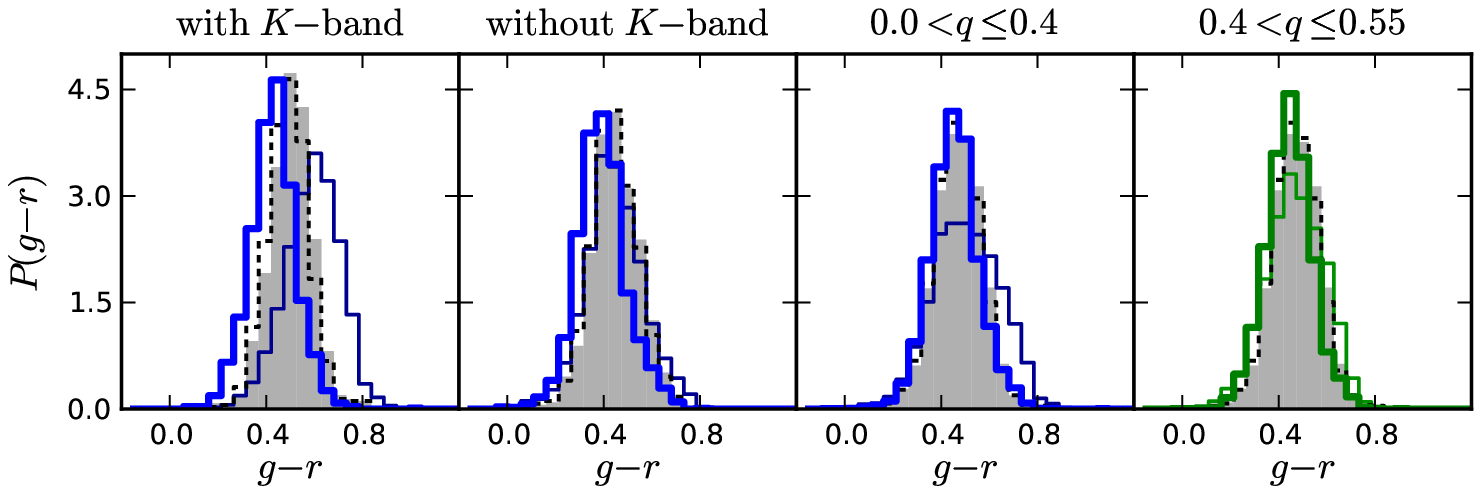}
    \caption{Corrected colour distributions for galaxies fit with a
      single exponenital profile. In all panels, the shaded region
      shows the uncorrected colour distribution for face-on ($q_d >
      0.85$) galaxies, the black dotted line shows the corrected colour
      distribution for face-on galaxies. In the first three panels,
      the thin and thick blue lines show the uncorrected and
      corrected colours for edge-on ($q_d<0.4$) galaxies. In the last
      panel, intermediate inclination galaxies ($0.4<q_d\leq 0.55$)
      are shown. The first two panels show the M09 correction
      applied to galaxies with and without $K$ band data. The final
      panels show all the disk galaxies using the M09 correction plus
      our adjustment for the over correction.}
    \label{fig:maller_exp}
  \end{figure*}
}

\newcommand{\figMallerBDa}{
 \begin{figure}
    \includegraphics[width=3.2in]{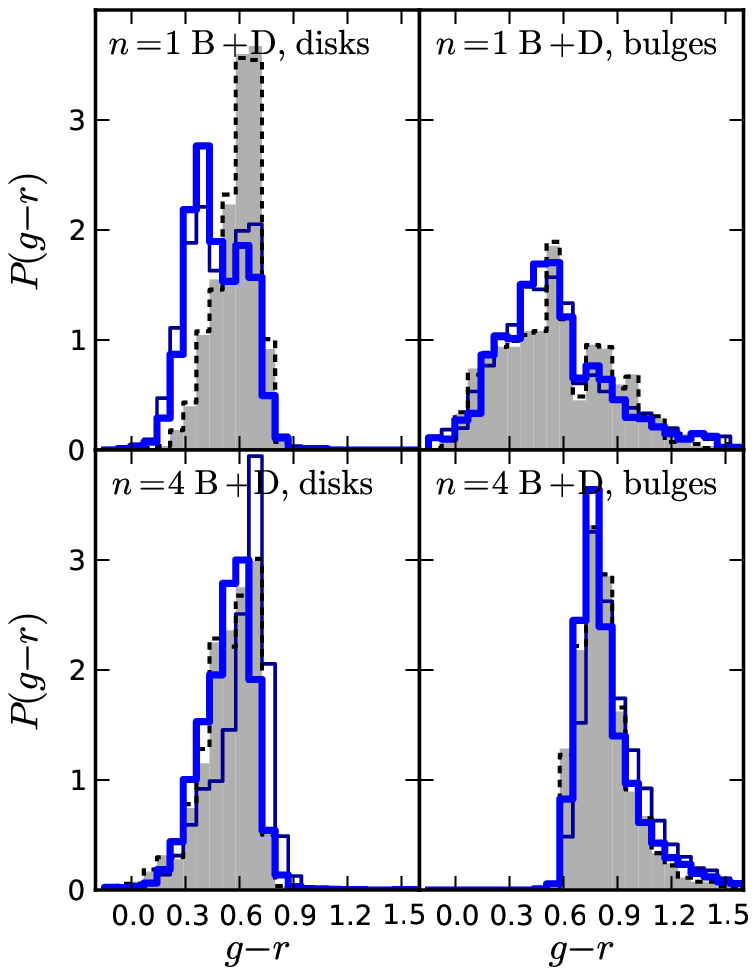}
    \caption{Corrected and uncorrected colour distributions for
      galaxies fit with $n_b=1$ B+D models (top) and $n_b=4$ B+D models
      (bottom). The thick and thin blue lines show corrected and
      uncorreted distribtuions for galaxies with $q_d < 0.4$. The
      shaded region and the black dotted lines show the distributions of
      uncorrected and corrected colours, respectively, for galaxies with $q_d >
      0.85$.}
    \label{fig:maller_BDa}
  \end{figure}
}

\newcommand{\figMallerBDb}{
 \begin{figure}
    \includegraphics[width=3.2in]{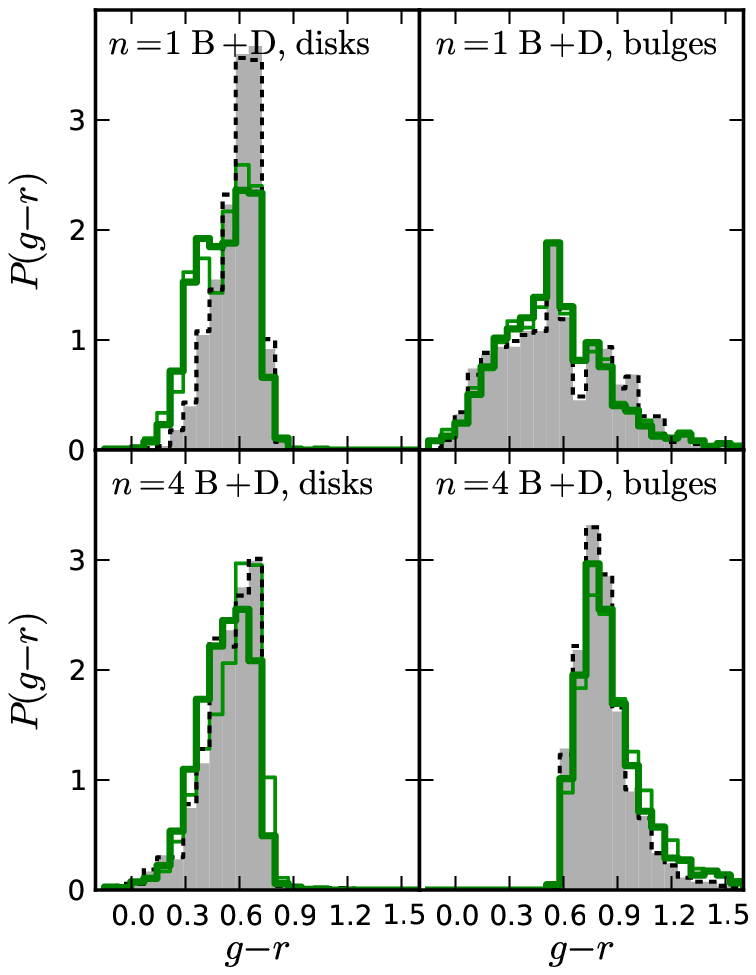}
    \caption{Same as Figure \ref{fig:maller_BDa}, but for galaxies
      with intermediate inclinations ($0.4<q_d<0.55$).}
    \label{fig:maller_BDb}
  \end{figure}
}

\newcommand{\figAGNCompare}{
 \begin{figure}
    \includegraphics[width=3in]{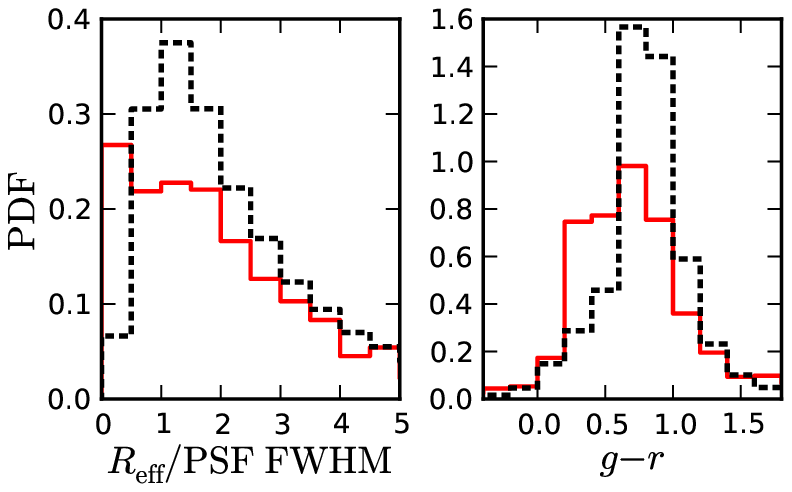}
    \caption{Probability distributions for the $n_b=4$ bulge \Reff{}
      and $g-r$ colour for all $B+D$ modeled galaxies (black
      dotted line) and the $1134$ BLAGN from \citet{Hao05a} (red solid
      line) fit with $n_b=4$ B+D models. The AGN sample is $1.3\%$ of
      the whole sample.}
    \label{fig:agn_compare}
  \end{figure}
}

\newcommand{\figBTTConc}{
 \begin{figure}
    \includegraphics[width=3.in]{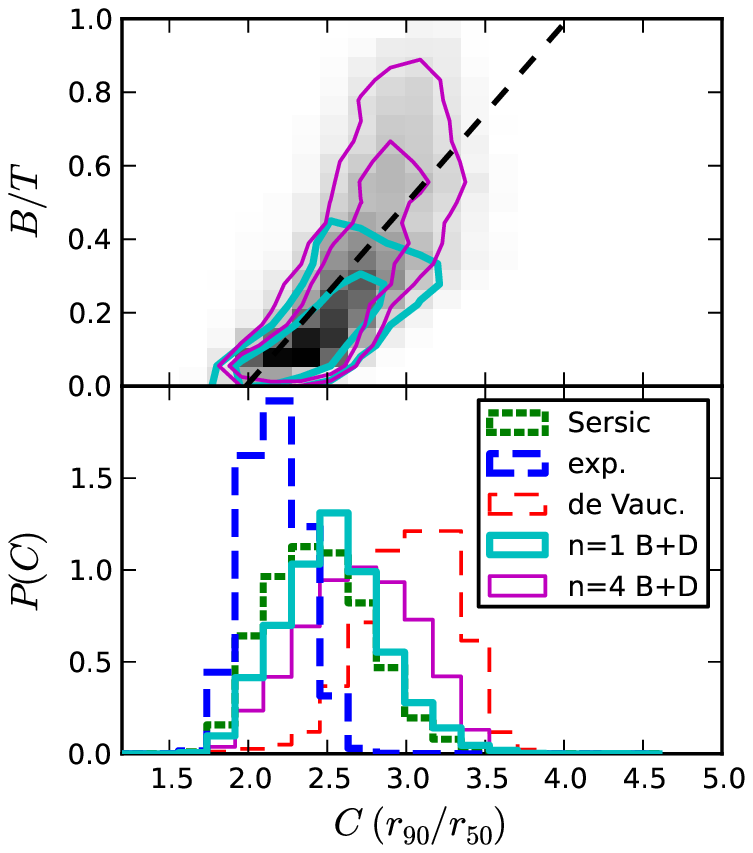}
    \caption{(\em{Top}\em): Relation between $B/T$ and Petrosian
      concentration for the sample of 
      $20,726$ B+D fitted galaxies. The underlying grayscale map includes
      all galaxies, the thin magenta (thick cyan) contours show the
      $n_b=4(1)$ B+D 
      modeled galaxies. The two contours in each colour enclose $50\%$ and
      $80\%$ of each sample. (\em{Bottom}\em): Distributions
      of Petrosian concentration for the different categories of galaxies. }
    \label{fig:BTT_Conc}
  \end{figure}
}
\newcommand{\figBTTN}{
 \begin{figure}
    \includegraphics[width=3.in]{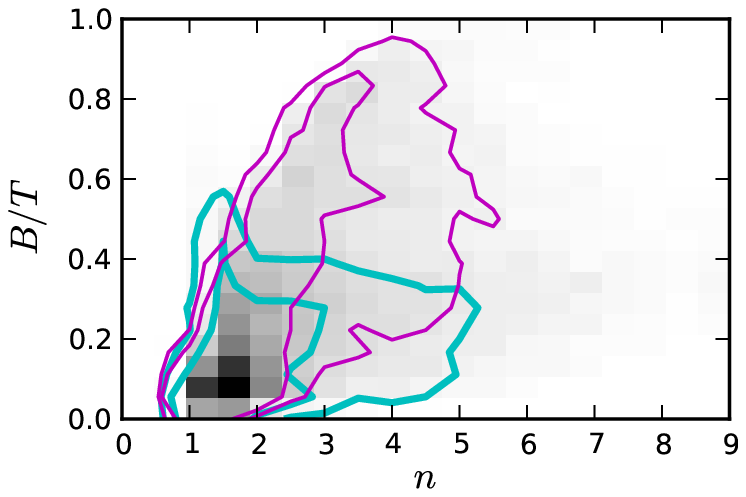}
    \caption{Relation between $B/T$ and the S\'ersic index of a single
      component fit for the B+D fitted galaxies. The symbols are the
      same as in the top panel of Figure \ref{fig:BTT_Conc}. }
    \label{fig:BTT_N}
  \end{figure}
}

\newcommand{\figCMDoverlay}{
 \begin{figure}
    \includegraphics[width=3.in]{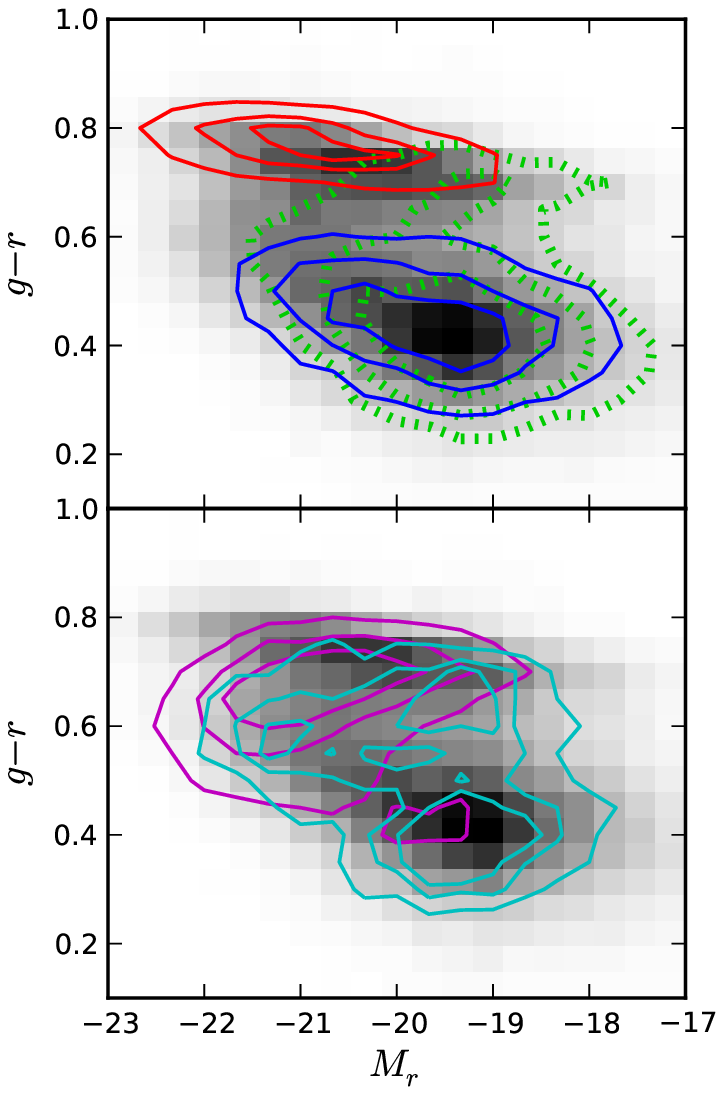}
    \caption{Colour magnitude diagram for different categories in the
      sample. The underlying grayscale is for the whole sample, using
      the assigned category to derive magnitudes and colours. In the top
      panel, the blue contours are the galaxies fit with a single
      exponential disk and the red contours are the galaxies fit
      with a single de Vaucouleurs profile. The green, dotted contours
      show the galaxies fit with a single S\'ersic profile. In the
      lower panel, the cyan contours are the $n_b=1$ B+D galaxies and
      the magenta contours are the $n_b=4$ B+D galaxies. The contours enclose
      $25\%$, $50\%$, and $75\%$ of each category. The colours and
      magnitudes are corrected using the corrections in
      \S\ref{ssec:inclination}.} 
    \label{fig:CMD_overlay}
  \end{figure}
}
\newcommand{\figCMDoverlayBD}{
 \begin{figure}
    \includegraphics[width=3.in]{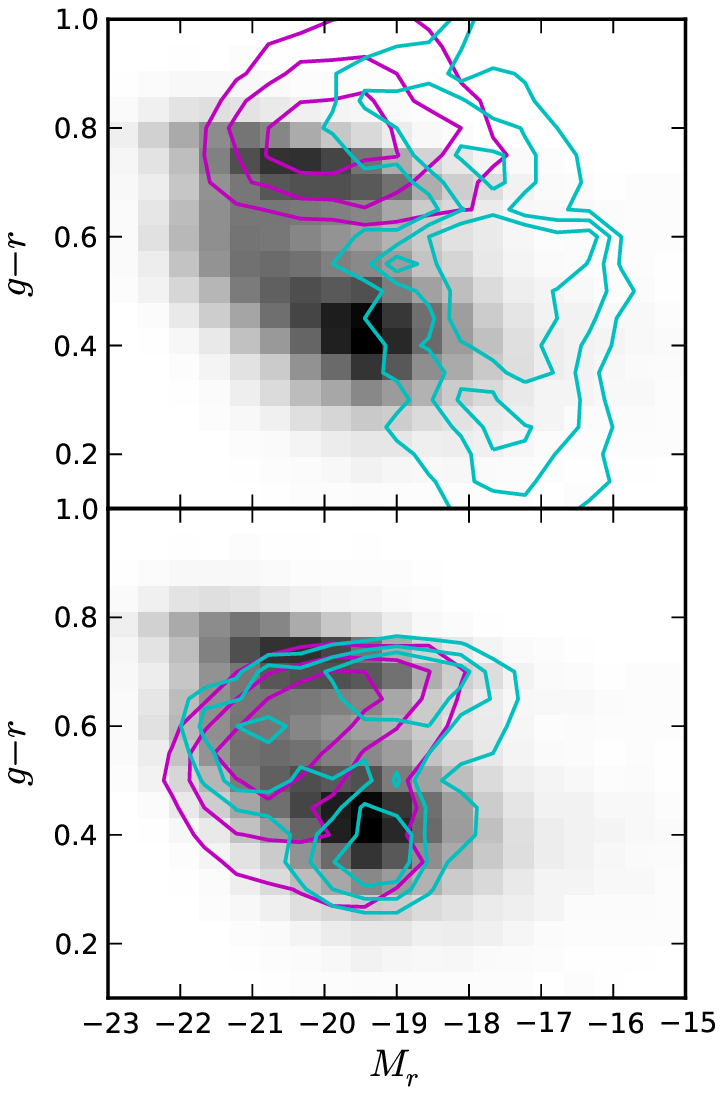}
    \caption{Colour magnitude diagram for bulges(top) and
      disks(bottom). The underlying grayscale is the CMD for the whole
      sample; the bulges and disks have been corrected for
      inclination. The cyan countours are the $n_b=1$ B+D galaxies, and
      the magenta countours are the $n_b=4$ B+D galaxies. The contours
      enclose $25\%$, $50\%$, and $75\%$ of each category. }
    \label{fig:CMD_overlayB}
  \end{figure}
}

\newcommand{\figCMDGreenValley}{
  \begin{figure}
    \includegraphics[width=3.2in]{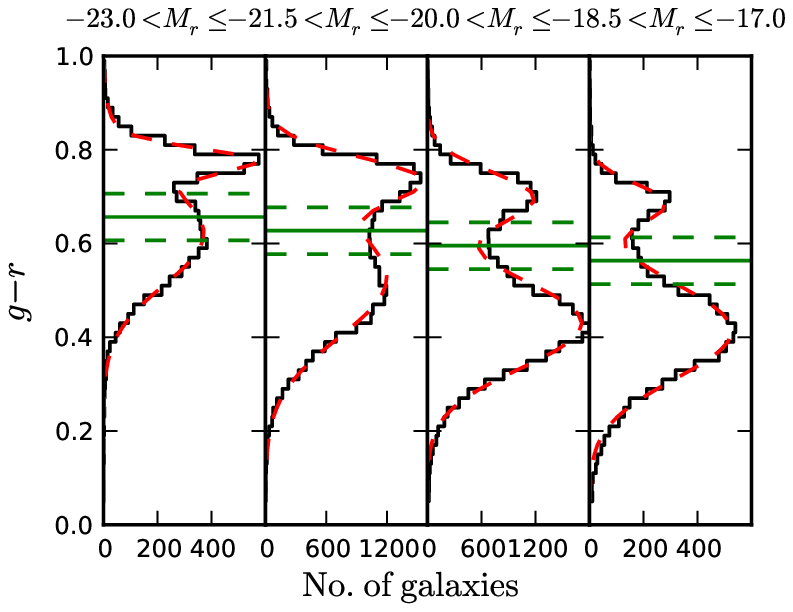}
    \caption{Colour histograms binned by absolute magnitude. The red
      dashed curve shows a double Gaussian fit to the histogram. The
      green solid and green dashed lines show the center and limits of
      the green valley as defined in \S\ref{ssec:bgr_gals}. }
    \label{fig:CMD_greenvalley}
    \end{figure}
}

\newcommand{\figbgrBDFits}{
  \begin{figure*}
    \includegraphics[width=6in]{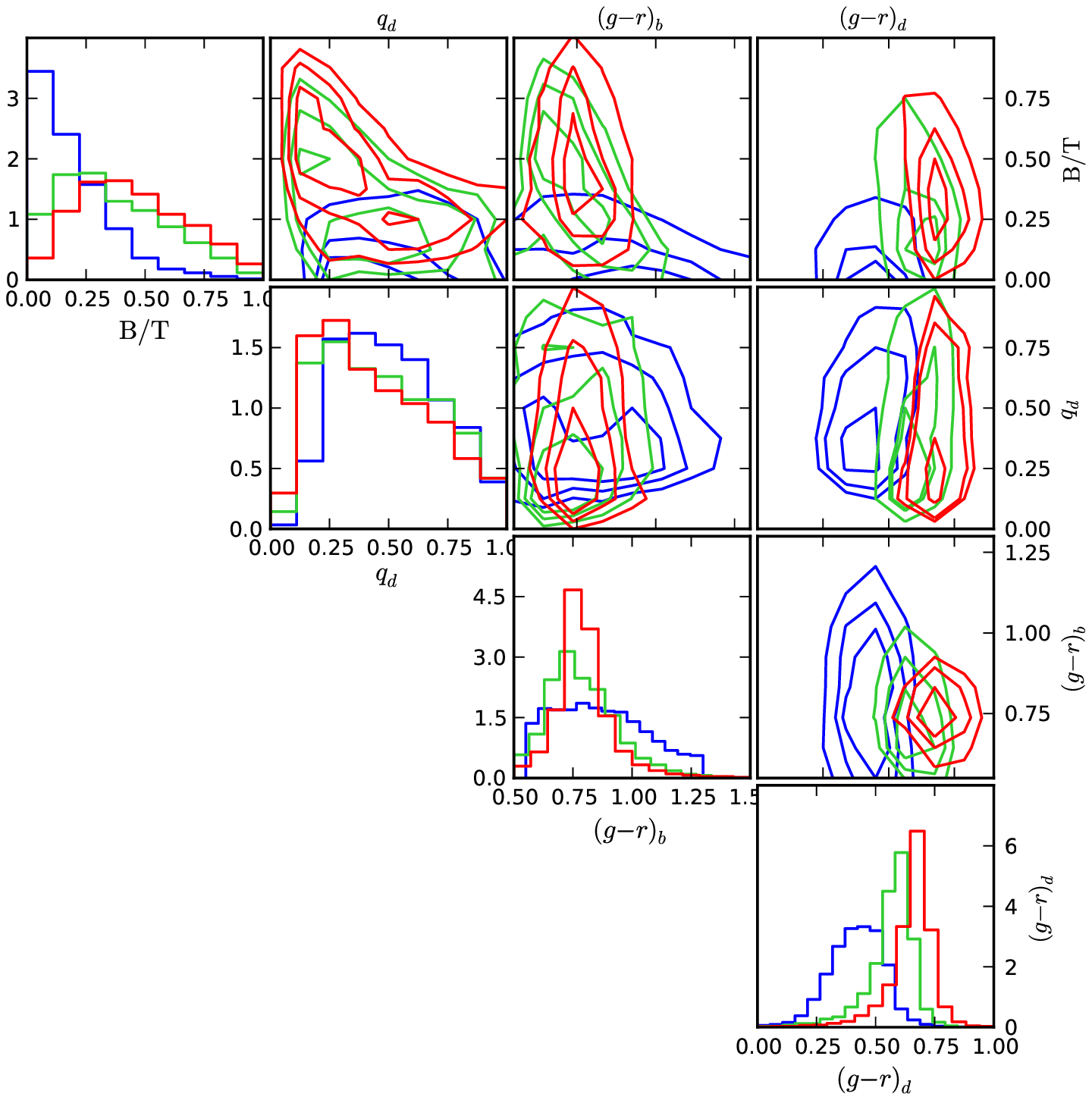}
    \caption{Properties of blue, green, and red B+D galaxies. The
      columns(rows) are: $B/T$, disk axis ratio, disk colour, and
      bulge colour. All quantities are corrected for
      inclination. The histograms are normalised to aid in the
      comparison of the different 
      colours. There are $8657$ blue galaxies, $5899$ green, and $5956$
      red galaxies. The galaxies are modeled with an $n_b=4$ B+D or an
      $n_b=1$ B+D according to the criteria in
      \S\ref{ssec:n1n4bulge}. However, the results do not change if
      only $n_b=1$ or only $n_b=4$ B+D models are used.}
    \label{fig:bgr_BD_fits}
  \end{figure*}
}

\newcommand{\tabGcomp}{
  \begin{table}
    \label{table:G09comp}
    \caption{Table comparing G09 categorizations to this work. The rows show the G09 categorizations and the columns show this works categorizations.}
    \begin{tabular}{lccccr}
      \hline
      & classical & pseudo & elliptical & no bulge & total \\
      \hline
      $n_b=4$ B+D & 46  & 12 & 11 & 0 & 69 \\
      $n_b=1$ B+D & 39  & 15 & 1  & 0 & 55 \\
      de Vauc.    & 103 & 2  & 84 & 0 & 189 \\
      exp. disk   & 5   & 42 & 0  & 4 & 51 \\
      S\'ersic    & 22  & 45 & 4  & 7 & 78 \\
      \hline
      total       & 215 &116 & 100& 11& 442 \\
    \end{tabular}
  \end{table}
}

\title[Bulge-Disk Decomposition of SDSS Galaxies]{Astrophysically Motivated Bulge-Disk Decompositions of SDSS Galaxies}

\author[C. N. Lackner and J. E. Gunn]{C. N. Lackner$^{1}$\thanks{E-mail:
clackner@astro.princeton.edu} and J. E.
Gunn$^{1}$ \\
$^{1}$Department of Astrophysical Sciences, Princeton University, Princeton, NJ
08544}

\begin{document}
\date{\today}
\pagerange{\pageref{firstpage}--\pageref{lastpage}} \pubyear{2011}
\maketitle

\label{firstpage}

\begin{abstract}
We present a set of bulge-disk decompositions for a sample of $71,825$ 
SDSS main-sample galaxies in the redshift range $0.003<z<0.05$. We
have fit each galaxy with either a de Vaucouleurs (`classical') or an
exponential (`pseudo-') bulge and an exponential disk. Two dimensional
S\'ersic fits are performed when the 2-component fits are not
statistically significant or when the fits are poor, even in the
presence of high signal-to-noise. We study the robustness of our
2-component fits by studying a bright subsample of galaxies and we
study the systematics of these fits with decreasing resolution and
S/N. Only $30\%$ of our sample have been fit with two-component fits
in which both components are non-zero. The $g-r$ and $g-i$ colours of
each component for the two-component models are determined using
linear templates derived from the $r$-band model. We attempt a
physical classification of types of fits into disk galaxies,
pseudo-bulges, classical bulges, and ellipticals. Our classification
of galaxies agrees well with previous large B$+$D decomposed
samples. Using our galaxy classifications, we find that Petrosian
concentration is a good indicator of B/T, while overall S\'ersic index is
not. Additionally, we find that the majority of green valley galaxies
are bulge$+$disk galaxies. Furthermore, in the transition from green
to red B+D galaxies, the total galaxy colour is most strongly
correlated with the disk colour.  
\end{abstract} 


%
\section{Introduction}
\label{sec:intro}

The division of galaxies into spirals and ellipticals and the
identification of the central bulges of spirals with ellipticals is
very old \citep{Hubble1936, dVc59, Sandage61}. Reasonably successful
quantitative attempts to fit galaxies with elliptical-like bulges and
exponential disks \citep{dVc48,dVc59,Freeman70} are almost as old as
the availability of imaging 
detectors \citep{Kent85,Kent86,Capaccioli1987}. It is not yet clear
what evolutionary forces give rise to the two components nor is it
clear quantitatively how general the division of bulge and disk
is. For example, it has long been recognized that low-luminosity
star-forming systems often have no bright central structure (bulge) at 
all. More recently, it has been recognized that the bright central
regions of (particularly) late-type spirals are often {\it not}
`classical' de Vaucouleurs-like bulges, but flattened subsystems \citep{Kormendy93,Fathi03}, sometimes 
with their own spiral structure, undergoing continuing star
formation, and with brightness profiles like those of disks
\citep{Kormendy77, Kormendy93, Fisher08}. These bulges have been
coined pseudo-bulges \citep[see][for complete set of pseudo-bulge
characteristics]{Kormendy04} and are thought to originate from secular
processes in disks (bars, spiral structure, etc.)
 \citep{Kormendy04, Athanassoula05, Weinzirl09}. On the 
other hand, `classical' de Vaucouleurs-like bulges, are thought to have
formed early by hierarchical mergers \citep{ELS62}, and acquired their disks
later \citep[but see][]{Noguchi99,Elmegreen08}.

For galaxies with classical bulges, the differences between bulge and
disk components extend beyond morphology. The stellar populations,
mean ages, dynamics, and star formation rates are very different,
again suggesting that bulges are very much like ellipticals, which
also have old stellar populations, pressure-dominated dynamics, and
low star-formation rates. Like ellipticals, bulges have old stellar populations and abundance ratios consistent with an early burst of star
formation \citep{Peletier99, Moorthy06, MacArthur08}. Additionally,
studies have found that classical bulges lie on an extension of the
fundamental plane \citep{FalconBarroso02}\citep[but see][]{Laurikainen2010}, suggesting similar dynamics. The (typically) much higher star formation
rates in disks give rise to significant colour differences between
bulges and disks.  This blue disk/red bulge phenomenon is clearly a
major reason for the observed colour bi-modality in galaxies
 \citep{Driver06,Drory07}. However, it is also clear that the real situation
is complicated and environment-dependent, with the present-day
universe containing both blue elliptical galaxies with ongoing star
formation \citep{Schawinski2009}, and galaxies with prominent disks and
essentially no star formation activity \citep{Bamford09, vandenBergh09,
Masters2010}.

The existence of the data set from the Sloan Digital Sky Survey (SDSS)
makes it possible to investigate the properties of bulges and disks
for a very large sample. We present here two-dimensional fits,
including the colours for bulge and disks components, for
71,825 main-sample SDSS galaxies with redshifts between 0.003 and
0.05. We attempt
from the outset to obtain fits which are astrophysically motivated and
which make morphological sense. The population is so complex that this
is only partially successful, as we show below. Guided by the division
of bulges into classical and 
pseudo-bulges, we fit either a de Vaucouleurs or exponential bulge and
an exponential disk to each galaxy for which a two-component fit makes
statistical sense. For the galaxies with signal-to-noise too low to
significantly support a two-component fit {\it and} those for which
the bulge plus disk (B+D) fit is physically implausible, despite high
S/N, we fit elliptical S\'ersic profiles (for which the surface
brightness falls as $R^{1/n}$) \citep{Sersic69}.  It is clear from our
work that generalizing fits to allow more parameters is not justified,
at least for this data set, and studies of bright subsets of our
sample suggest that it may not be useful for ground-based images at all. 

Both qualitative and quantitative methods for determining
galaxy morphology are used. For nearby, well-resolved
samples, two dimensional decomposition of galaxy images into bulge and
disk components is possible and there are several automated tools
available to do such decompositions: \verb+GALFIT+ \citep{Peng02},
\verb+GIM2d+ \citep{Simard02}, \verb+BUDDA+ \citep{deSouza04,Gadotti08}, and 
\verb+GALACTICA+ \citep{Benson07}. These methods have been used to
create large samples of bulge-disk decomposed galaxies  \citep{Allen06,
  Benson07,Gadotti09, Simard2011}. With the exception of
\citet{Simard2011}(S11), the largest sample consists of $10,000$ galaxies from
the SDSS early data release \citep{Benson07}. The sample of
S11 includes $1.1$ million galaxies from the SDSS
Legacy survey using data release 7. This sample includes $96\%$ of the
galaxies in the sample presented here; the missing galaxies are
brighter than the cutoff in the S11 work. Nonetheless,
the numerical methods used in the works are distinct, and the goals of
the two studies are not the same. In the future,
we plan to use the sample presented in this work to study the properties of disks around
classical bulges, and have therefore focused on robust fits to
galaxies with statistically significant bulges and disks. 

 In order to accommodate this large sample, we have written a
 bulge-disk decomposition pipeline in \verb+IDL+ which relies heavily
 on the SDSS reductions. We use the SDSS sky subtraction, deblending,
 and point-spread function (PSF) determinations in our fitter. This
 allows us to quickly fit each galaxy in our sample with $5$ different
 models, using a uniform methodology. In addition to fast fitting
 algorithms, large samples of bulge-disk decomposition require careful
 analysis of the systematic errors in estimating bulge and disk
 components as a function of galaxy size, luminosity, and redshift. At
 the extremes, for galaxies with sizes comparable to the PSF, a
 bulge-disk decomposition does not yield meaningful
 results. Similarly, images of nearby galaxies contain structure which
 is not accounted for in simple two-component bulge-disk models,
 yielding poor quality, inaccurate fits. We have examined the
 systematic errors in our sample by fitting galaxy images while
 varying the signal-to-noise and resolution of galaxy images, as well
 as degrading images of nearby galaxies to match those at higher
 redshift. These tests give an estimate of the systematic and
 statistical uncertainty in the bulge-disk parameters for the galaxies
 in our sample. 

Due to the size of our sample, we must rely on output of the
bulge-disk decomposition to classify the galaxy as an elliptical or
disk galaxy, and the bulge as a classical bulge or pseudo-bulge. At
high resolution, galaxies are almost always more complex than 
either of our B+D models. For lower resolution and
signal-to-noise images, the two models are often indistinguishable,
fitting 
the galaxy equally well. Therefore, in addition to a goodness of fit,
we must on astrophysical constraints in order to quantify the
galaxy morphology. Since we expect classical bulges to be
elliptical-like and pseudo-bulges to be disk-like, we distinguish
between  classical and pseudo-bulges based
on their colour and ellipticity. Blue,
flattened bulges are categorized as
pseudo-bulges and assigned an exponential light profile, while red
bulges are assigned a de Vaucouleurs profile. We test this division of
galaxy type against other measures of galaxy type (e.g. S\'ersic index,
concentration, fundamental plane relations) and find that the
combination of colour information with bulge-disk decomposition yields
a reasonable classification of galaxy types. Although
we have fit more than $70,000$ galaxies with two-component fits, we do
not expect these fits to be physically meaningful in all
cases. For $50\%$ of our sample we do not attempt to classify bulges
as a classical or pseudo, instead, we fit these galaxies with single
S\'ersic profiles, and set them aside. Some of these galaxies
undoubtedly have a bulge and disk, but are not bright enough or
sufficiently resolved to support a good B+D fit. However, we find that
the majority of galaxies fit with a S\'ersic profile are intrinsically
faint and blue, and are therefore unlikely to be well-fit by a B+D
model.

This paper is divided into sections as follows. In \S\ref{sec:sample} we
describe the sample of galaxies. Section \ref{sec:fitter} describes the
fitting procedure and \S\ref{sec:qualfits} examines the robustness of
the fits by examining the impact of changes in signal-to-noise,
resolution and redshift. In section \S\ref{sec:analysis}, we divide
our sample into pure disks, ellipticals, disks plus classical bulges
and disks plus pseudo-bulges, taking into account inclination
corrections (\S\ref{ssec:inclination}). Throughout this paper we use
the $\Lambda\mathrm{CDM}$ cosmology, $\Omega_m = 0.3$, $H_0 =
70\kmps$, and $\Omega_\lambda = 0.7$. 

%
\section{Sample}
\label{sec:sample}
The sample of galaxies we fit is a low redshift sub-sample of the NYU
value-added catalog (VAGC) of the SDSS Data Release 7 (DR7) galaxies
 \citep{Blanton2005, SDSSDR72009, Padmanabhan08}. We have matched the
DR7 catalog to objects and images from DR8  \citep{SDSSDR82011}, and
use the  later data release images for our analysis. The most
important difference between the DR8 and DR7 reductions is the
improved sky subtraction around bright, extended objects \citep{SDSSDR82011}. All
the galaxies in our sample are also contained in the SDSS
spectroscopic sample. We limit the redshift range of the sample to
galaxies with $0.003 < z < 0.05$ and magnitudes $m_r < 17.7$. We
further remove all galaxies with poor deblends flagged by the SDSS
photometric pipeline. The galaxy fluxes were k-corrected to $z=0$ using the
\verb+IDL+ package \verb+kcorrect+, \verb+v4_2+ \citep{Blanton07}. At
this point, the sample consists of $87,000$ galaxies. 

Because we have chosen
to fit simple S\'ersic profiles to the galaxies, our models are
inaccurate for highly-inclined disk galaxies, where the vertical scale
height is significant relative to the minor axis of the projected
galaxy. Studies of edge-on disk galaxies have shown that the minimum
disk flattening (vertical scale height/disk scale length) is $\sim 5$
 \citep{Kregel02}, and is independent of galaxy size. However, the
\emph{average} disk flattening increases with the maximum rotation
velocity of the disk. Flat disk models used to fit highly-inclined
galaxies will introduce errors in the surface brightness and the
inclination. We have examined the size of these errors as a function
of inclination for projected three-dimensional disk galaxy models. We
model a disk galaxy as exponential profiles in $R$ and $z$ with a disk
flattening of $5$, and then fit projections of this model with a
flat disk (exponential profile in $R$) model. When the inclination of the disk
is more than $\sim 75^\circ$, (corresponding to an axis ratio of $0.25$
for a flat disk), the errors in the measured axis ratio and surface
brightness are between $5\%$ and $10\%$. In order to limit the errors
due to poor disk models, we have limited our work to galaxies with
axis ratios greater than $0.25$, as measured by the SDSS
pipeline. This leaves a sample of $79,476$ galaxies. 

The images used in the fitter are the SDSS generated atlas images, which
contain all the pixels around a source with significant
detection. S11 argue that the isophotal cutoff for the
atlas images is too bright, leading to significant loss of flux. We
can check this by calculating how much of the fitted model flux is
outside the atlas image for each galaxy. For single S\'ersic profile fits,
between eighty and ninety percent of the atlas images contain $90\%$
of the model flux. The uncertainty is due to not knowing how many
pixels are excluded from the atlas images because they are part of
other objects. The bulge$+$disk models are missing a comparable amount
of light. Unlike the models used in S11, our
models are cutoff at large radii, reducing the amount of light missing
from the atlas images in our model fits.

The atlas images are deblended and sky subtracted. The
value of the sky is the locally measured SDSS sky value, not that of
the entire field. The SDSS photometric pipeline, \verb+photo+, 
calculates the sky on a grid of $128\times 128$ pixels by taking the clipped median of
$256\times256$ pixels surrounding each grid point. The local sky is
obtained by bi-linear interpolation. The new version of \verb+photo+
used in DR8 first subtracts preliminary models (linear combination of
de Vaucouleurs and exponential) for the bright galaxies and then takes
the median around each grid point \citep{SDSSDR82011}. This helps
eliminate flux from bright extended objects from the estimate of the
sky, a known problem with earlier SDSS reductions
\citep{Mandelbaum2005,Bernardi07,Lauer07,Guo2009,SDSSDR72009}. The
preliminary galaxy models are added back in and refit after 
the sky subtraction is complete.

\section{Fitting}
\label{sec:fitter}
We perform the 2-dimensional fits on the $r$-band atlas images. Each
image is fit five times, using five different profiles, two composite
(bulge+disk) profiles and three single component profiles: 
\begin{enumerate}
\item de Vaucouleurs bulge + exponential disk profile
\item exponential bulge + exponential disk profile
\item S\'ersic profile
\item exponential disk profile ($n=1$ S\'ersic profile)
\item de Vaucouleurs profile ($n=4$ S\'ersic profile)
\end{enumerate}
The functional form of a single S\'ersic component is given by the formulae:
\begin{eqnarray}
\label{eq:fits}
S&=&\Sigma_{1/2} \exp\left[-k\left(\left(R/\Reff\right)^{1/n}-1\right)\right] \\
R&=&\surd\left[\left(\left(x-x_0\right)\cos\phi+\left(y-y_0\right)\sin\phi\right)^2\right.\\	
&+&\left.\left(\left(y-y_0\right)\cos\phi-\left(x-x_0\right)\sin\phi\right)^2/q^2\right]\quad,\nonumber
\end{eqnarray}
where \Reff{} is the half-light radius of the profile,
$\Sigma_{1/2}$ is the surface brightness at \Reff, $(x_0,y_0)$ is the
central position, $\phi$ is the rotation angle, $q$ is the 
axis ratio of the elliptical isophotes, $n$ is the S\'ersic index, and
$k$ is a normalization factor given by $k = n \times \exp \left( 0.6950 -
  0.1789/ n \right)$ \citep{LimaNeto99}. Therefore, for the single
component there are 6(7) free parameters for the de 
Vaucouleurs and exponential disk profiles (S\'ersic profile). For the
two component fits, we assume the central position is the same for the
bulge and the disk, leaving $10$ free parameters for the bulge/disk
fits. As with the SDSS de Vaucouleurs and exponential profiles, our
model profiles are cut off smoothly at large radii. For
exponential (de Vaucouleurs) profiles, the surface brightness is
suppressed outside 3(7)\Reff{} and drops to zero outside 4(8)\Reff.

We have chosen not to allow the S\'ersic index of the bulge be a
free parameter in the fits. This is done to limit the number of free
parameters in the fits. The bulge S\'ersic index
is highly covariant with the half-light radius of the bulge
 \citep{Trujillo01}, and for a typical galaxy in our sample, varying
the S\'ersic index does 
not lead to better fits \citep[][see also]{deJong1996}.  As can be seen
in equation \ref{eq:fits}, the differences between different S\'ersic
index profiles decreases with increasing S\'ersic index. High S\'ersic
index profiles are sharply peaked and decrease slowly outside
\Reff. For bulges embedded in disks, the long tail is subsumed 
in a disk, while the central peak is washed out by the
PSF. At a resolution typical for our sample, for a
galaxy with $m_r\approx16.5$, a bulge with $n_b=4$ and a $B/T \approx
40\%$, the $\chi^2$ value of a fit with an $n_b=4$ bulge is
indistinguishable from one with $3 \la n_b \la 5$. However, for this
range of S\'ersic index, \Reff{} varies by a factor of $2$, which
leads to significant variation in the $B/T$ \citep{Graham01,
  Balcells03,Graham08}. 
In order to
facilitate a comparison with  elliptical galaxies, we have chosen to
fit classical bulges and ellipticals with a de Vaucouleurs
profile. This value for $n$ may be too large for bulges (and for
some ellipticals \citep{Caon93}), but it eliminates a parameter which does
little to improve the $\chi^2$ values of our fits, and allows us to
compare bulge properties to those of ellipticals.  Because we
anticipate our sample will contain pseudo-bulges and bulges of lower
S\'ersic index, we have also fit each galaxy with an exponential
($n_b=1$) bulge + exponential disk. In the latter case, we define the
bulge component to be the profile with the smaller \Reff{}. These two
bulge S\'ersic indices are in broad agreement with the
S\'ersic indices for pseudo-bulges ($n_b\approx 1.7$) and classical
bulges ($n_b \approx 3.5$) found in smaller, high resolution (HST) studies
\citep{Fisher08}. 

The best-fitting parameters for each model are found using the
minimization method \verb+mpfit2dfun+ in 
IDL, which performs 2-dimensional Levenberg-Marquadt minimization
\citep{Markwardt09}. The fitter minimizes the weighted sum of the
difference between the image and PSF-convolved model. The fitted images are
convolved with the locally measured PSF from
SDSS. The weight of each pixel is given by its inverse variance. The
variance in counts is given by $($signal$+$sky$)/$gain$+$dark
variance$+$sky error$^2$. For sky-dominated images, the variance is
constant as a function of galaxy radius. For bright galaxies, the
variance is larger, and the weights smaller, at the centres of the
galaxies. Although the inverse variance weighting are the optimal
weights for a $\chi^2$ fit, they introduce systematics with redshift
(and size and brightness): i.e. galaxies at high redshift are fit with
nearly constant weights, while bright, nearby galaxies are
down-weighted in the centres. For samples covering a larger redshift 
range, a set of weights which are consistent across the range may be
more appropriate. This would entail using less than optimal weights
(e.g. using only the sky-noise for the weights) for the highest
resolution and brightest galaxies. 

 The initial parameters for the minimization are taken from
the SDSS analysis of the galaxy. The initial fit parameters include  
the one dimensional de Vaucouleurs or exponential fit scale length,
chosen based on \verb+FRAC_DEV+, the relative likelihood of a
de Vaucouleurs fit over and exponential fit,
\verb+R_(DEV,EXP)+, the angle of rotation of profile,
\verb+PHI_(DEV,EXP),+ the axis ratio of either profile,
\verb+AB_(DEV,EXP)+. We require that the fitter return positive values
(but possibly zero) for the surface brightness for each component. We also
place lower limits on \Reff{} (0.1 pixels) and on $q$ (0.05). We allow
the minimizer $1000$ steps. Fits that do not converge within $1000$
steps are considered failed fits. Fits in which either bulge+disk
combination fail to converge occur for $2.5\%$ of the sample. Fits in which both
B+D models fail occur in less than 20 images in the entire
sample. We also exclude galaxies in which the signal-to-noise within
$\Reff$ is less than $3$, for any of the single component fits. This typically
excludes diffuse, irregular galaxies for which none of the above
models is reasonable. This excludes $3530$ galaxies from the sample. 

In order to examine the colours of the components of each galaxy, we
also fit the $g$ and $i$-band atlas images of each galaxy. These fits
are accomplished by a taking the best-fitting $r$-band model, and
performing a linear least squares for the $g$ and $i$-band images by
scaling the surface brightness, $\Sigma_{1/2}$, for each profile
component. Therefore, the profiles fit in each band are the same, only
the overall normalizations and the bulge-to-total ratios ($B/T$) change
in the different bands. This procedure is analogous to the SDSS \verb+modelMag+
measurements, which apply the best $r$-band model (de Vaucouleurs or
exponential) to all the filters. These aperture-matched colours are
less noisy than those that would be obtained by fitting a B+D model
independently for each filter. 

We have compared our single component exponential and de Vaucouleurs
fits to those from SDSS. Overall, we find good agreement between our
fits and those from SDSS when the fit is accurate for a galaxy,
i.e. when \verb+FRAC_DEV+ is large(small) for de 
Vaucouleurs(exponential) fits. For the exponential fits, the median
difference in $r$-band magnitude is $0.040$ with an inter-quartile
range (IQR) of [$0.018$, $0.072$], for all the
galaxies. Those with the largest offset are galaxies which have high
\verb+FRAC_DEV+, and are poorly fit by an exponential profile. For the
de Vaucouleurs model, the median difference in magnitude is
$-0.159$ (IQR=[$-0.381$, $-0.027$]). This difference decreases to
$-0.025$ (IQR=[$-0.065$,$0.017$]), if we restrict the comparison to galaxies
with \verb+FRAC_DEV+$>0.5$. This suggests that our fully $2$-dimensional
models fail differently than the SDSS single component models. The
cause of this is probably related to how the image is cut off at large
radii; de Vaucouleurs models are much more sensitive to extended flux
than exponential models, making the differences larger. For the
exponential fits, the half-light radii measured agree to within $\la
5\%$ for all fits and to $\la3\%$ for fits with
\verb+FRAC_DEV+$<0.5$. For the de Vaucouleurs \Reff{}, the
difference in \Reff{} is larger, with a median difference between
the SDSS \Reff{} and our \Reff{} of $-9\%$, for 
galaxies with \verb+FRAC_DEV+$>0.5$. This difference is the main
reason our magnitudes are systematically brighter than the
SDSS de Vaucouleurs magnitudes. Furthermore, the fractional difference in scale
length increases with increasing \Reff.  The difference in scale
length has several causes. First, the SDSS fits use a maximum allowed
scale length of $80$ 
pixels while we allow \Reff{} to take any value larger than $0.01$
pixels. Second, the de Vaucouleurs profiles in SDSS are softened
within $\Reff/50$ in order to allow for a large dynamic range in surface
brightness. We do not use this softening, which leads to slightly
larger values for \Reff. The galaxies with the largest values for
\Reff{} are almost always better fit by a two-component or S\'ersic
model fit, so the large differences between our fits and those in SDSS
occur for galaxies for which the de Vaucouleurs model is inaccurate,
and therefore are not a large concern. 

We also compare our single component S\'ersic fits to those found by
\citet{Blanton2005a}, which use the azimuthally-averaged radial
profiles from the SDSS \verb+photo+ pipeline. We find systematically
higher S\'ersic 
indices for $n \ga 3$. This difference arises from the following
sources. We allow S\'ersic indices to vary between $0.1$ and $9$ while the 
\citet{Blanton2005a} S\'ersic indices do not exceed $6$. Additionally,
\citet{Blanton2005a} find that their high S\'ersic indices are
underestimated by $\sim 0.5$ at $n=4$, and claim this is due to uncertainty
in the sky. Our measurements are done on DR8 images which have
improved sky estimates. Finally, the details in the cutoff radius are
not the same, which will effect the high S\'ersic index galaxies more
than the low S\'ersic index galaxies. The mismatch between our
measured S\'ersic indices and those from \citet{Blanton2005a} is
similar to the mismatch found by S11.

\subsection{Examples of fits}
\label{ssec:exfits}
Figure \ref{fig:example_fits} shows B+D
fits for four bright, well-fit galaxies, in order of increasing
$B/T$. The top two galaxies are shown with an $n_b=1$ bulge fit, while
the bottom two are shown with an $n_b=4$ fit. The model fit is chosen
based on the relative $\chi^2$ values of the different B+D fits. The
residuals are shown in the far right column. The 
$1-$dimensional profiles for these galaxies are shown 
in Figure \ref{fig:example_profs}. 

\figeximages
The last panel in Figure \ref{fig:example_fits} is an
elliptical galaxy; however the fit includes a large, low surface brightness
exponential component along with the $n=4$ component. The $\chi^2$
value of this fit is statistically significantly better than the fit to a pure
de Vaucouleurs profile. The $B/T$ ratio of this galaxy is 0.86. The
exponential component has a central surface brightness $> 27\mathrm{\
  mag\ arcsec}^{-1}$, and a signal-to-noise of 1.7 within
\Reff. If the bulge were subtracted from the image, the disk alone
would be undetectable. Because we ignore galaxies for which S/N for a
single component fit is less than $3$, we have also removed individual
components
from our B+D fits which have 
S/N$<3$. Therefore, SDSSJ011457.59+002660.8 is best fit by a single de
Vaucouleurs profile. Additionally, the $g-r$ and $g-i$ colours of
exponential and 
de Vaucouleurs components for this galaxy are the same to within
uncertainty. This suggests the exponential component is most likely 
not a disk, but simply a reflection of the fact that the galaxy is not
a perfect de Vaucouleurs profile. The best-fitting S\'ersic profile
for this galaxy has $n=4.4$, which helps account for the exponential
profile at large radii. These so-called fake disks are a
known problem for automated bulge-disk decompositions \citep{Allen06,
  Cameron09}. One solution is to set a maximum $B/T$ for two component
fits and fit everything above the cutoff with a single S\'ersic
model \citep{Allen06}. Additionally, these exponential components are
characterized by having similar ellipticity to their bulge as well as
similar colours, and can be successfully removed based on a combination
of these three characteristics. We discuss choosing between the five
different model fits in \S\ref{ssec:n1n4bulge}. Tables containing
all the models fit to each galaxy as well as our best-fitting model
for each galaxy are available for download.\footnote{See
  http://www.astro.princeton.edu/$\sim$clackner/home/research for the
  complete data tables.}
\figexprof

\section{Quality of fits}
\label{sec:qualfits}
In order to assess the quality of the fits, we have created two
sub-samples. The first, Sample A, is a hand-selected sample
of ~30 bright ($M_r > 13.5$) galaxies which span a range of $B/T$ and
$q$ values. These galaxies are well fit by either B+D model,
but many of them have extra structure, such as spiral arms. We use
this sample to test the effects of degrading signal-to-noise and the
resolution in our images. The second sub-sample, Sample B, is a sample
of $190$ galaxies which span the magnitude and redshift range of our
sample. The selection of this sample is random and not based on the
quality of the fits, but galaxies for which either B+D model failed to
converge have been removed. 

\subsection{Colours}
\label{sssec:qualcolours}
\figcolourBTT
One way to test the robustness of our B+D decompositions is to perform
the full 2-dimensional B+D decomposition in the $g$ and $i$ bands as
well as the $r$-band. We have done this for Sample B. This test allows
use to estimate 
the errors of the colours of the bulge and disk components as measured
from the linearly scaled $g$ and $i$ band fits. From the reduced $\chi^2$ values, a probability value for
accepting/rejecting the model can be calculated from
\begin{equation}
\label{eq:chisqCDF}
p = \int_0^{\chi^2}P_\chi(\nu,\chi)\qquad ,
\end{equation}
where $\nu$ is the number of degrees of freedom and $P_\chi$ is the
$\chi^2$-distribution for $\nu$ degrees of freedom. Unlike the reduced
$\chi^2$ values, the probabilities can be directly compared between
the linearly scaled fits and the full model fits. While the average
difference between $p$-values for the full model fit and the linearly
scaled fits is $5\%$ and $4\%$ in the $g$ and $i$-bands, respectively,
the median difference in p-values is $0$.

In addition to comparing the quality of the fits in the different
bands, we also compare the actual parameters obtained in $g$, $r$, and
$i$. We find no systematic trends in $B/T$ with either magnitude
or $B/T$, although there is significant scatter. Figure
\ref{fig:btt_colour_comp} compares the $B/T$ in $r$ measured using the
models fit in $g$, $r$, and $i$. The $B/T$ values for $g$ and $i$-band
models are obtained by linearly scaling those models to fit the
$r$-band image. Since we are comparing $B/T$ in one band, we expect
the differences to be consistent with zero. Indeed, for the $n_b=4$ 
fits, the median values of $(B/T_r-B/T_{g,i})/B/T_r$, where the
subscript refers to the model, are $0.02$ (IQR=[$-0.04$, $0.11$]) and
$0.01$ (IQR=[$-0.06$, $0.10$]). For the $n_b=1$ B+D fits, the same
quantities are $0.02$ (IQR=[$-0.04$, $0.08$]) and
$0.05$ (IQR=[$0.02$,$0.09$]). All but the last value is consistent
with zero. For the exponential bulge fits, the large scatter at higher
$B/T$ is due to difficulty in separating the bulge from the disk when
they have the same functional form. The uncertainty in the $B/T$
measured in the different bands is directly related to the uncertainty
in the colours of the different components. A difference in $B/T$ of
$7\%$ corresponds to a difference in bulge colour of $-0.07$. If
instead of comparing $B/T$ measured in one band, we compare the $B/T$
measured in different bands, we expect the differences to be
consistent with the colour differences between bulges and disks. Since
bulges are redder than disks, $B/T$ should be largest in the redder
filters. Looking at galaxies which are categorized as bulge+disk
galaxies (see \S\ref{ssec:n1n4bulge}), bulges are typically $0.11$
magnitudes redder than the total galaxy in $g-r$. Similarly, the
median value of $(B/T_g - B/T_r)/B/T_r=0.13$, where the subscript refers to
the filter $B/T$ is measured in. This agrees well with difference
between bulge and total galaxy colours.

Comparisons of the scale lengths measured in the different filters is
shown in Figure \ref{fig:reff_colour_comp}. The fit parameter with the largest 
scatter is the $n_b=4$ bulge \Reff. For galaxies with $r$-band measured 
$B/T > 0.1$, the scatter in the de Vaucouleurs bulge \Reff{} is
$\sim20\%$. The scatter in the exponential bulge \Reff{} is
$\sim15\%$. The scatter in 
the disk \Reff{} in all three filters is $\la 10\%$. The large scatter in the
de Vaucouleurs scale length is related to the steepness of the $n=4$
S\'ersic profile compared to an exponential profile. Small changes in
the measured central flux yield large changes in the measured \Reff. 
\figcolourReff
Figure \ref{fig:reff_colour_comp} shows the relative difference between
the $r$-band measured \Reff{} to those measured in the $g$ and $i$
images. There is no systematic offset in the values, although the
scatter in the bulge (disk) scale length measurements grows with
increasing (decreasing) $B/T$. Although the scatter in the \Reff{} is
consistent with no systematic difference in \Reff, the $g$-band models
tend to have slightly larger scale lengths than those in $r$ and
$i$. This is consistent with the observation that galaxies are bluer
on the outskirts \citep{Franx89}, thus the scale lengths in the blue bands should be
larger. The same phenomenon can be observed in the single
S\'ersic component fits. In this case, the scale lengths measured in
the different bands are consistent to within $5\%$, but the S\'ersic
indices measured in the bluer bands are consistently larger, which
increases the flux at larger radii, as expected. Offsets of a few
percent in the scale lengths are in agreement with the results of the
separate band-pass fits in S11.

\subsection{Signal-to-Noise and Resolution}
\label{ssec:SN_RES}
In this section, we explore the effects of degrading the
signal-to-noise (S/N) and resolution of our galaxy sample
images. Although, our sample spans a small range in redshift, it is
important to understand the systematics introduced by changing the
resolution and the S/N, and the limits on the resolution and S/N at
which our fits are no longer reliable. In order to examine these
effects, we have performed several tests. First, we have taken mock
images of single S\'ersic component galaxies at various S/Ns and
resolutions. Fits of these mock images are compared to the input
model.
Although mock galaxies offer some insight, we also take the $39$ high
resolution, high S/N galaxies in Sample A and 
create new images of these galaxies both at lower resolution and
lower S/N. We then run the fitter on these degraded images and compare
the results to the original fit of the galaxy. Finally, we examine the
combined effect of changing resolution and S/N by taking the images in
Sample A and degrading the S/N and resolution simultaneously to create
images of the same galaxies as they would look at higher redshift. We
define the signal-to-noise ratio for each pixel as the object flux
divided by the noise, the square root of the total flux (object
\emph{plus} sky) divided by the gain.

The mock images used are single S\'ersic component galaxies with \Reff{}
fixed at $10.6$, $15.2$, $21.0$, or $36.2$ pixels and a S\'ersic index
of $1$, $2$, $3$, or $4$. The surface brightness at $\Reff$ is the
same for all the galaxies, and the total magnitudes of the galaxies
vary from $12.7>M>13.9$, similar to the Sample A galaxies. The mock
images are convolved with a Gaussian PSF with FWHM=$2.9$ pixels, the
median from Sample A. We add noise to the images assuming a sky
background of $115$ counts/pixel, the median sky level from Sample
A. These are the initial mock images. We then create two sets of
degraded mock images, one in which the PSF FWHM is increased, 
degrading the resolution, and one in which the sky background is
increased, degrading the signal-to-noise. For the decreased resolution
images, we convolve the mock galaxies with a PSF with FWHM $2.9\times$($1.5$, $2$, $4$, $5$, $6$, $7$, and $10$). We do not change the sky
background, the galaxy properties, or the pixel size. For the
decreased S/N images, we hold everything constant except 
the sky background, which is increased by factors of $1.5^2$, $2.^2$,
$4^2$, $5^2$, $6^2$, $7^2$, and $10^2$. For sky-noise dominated
images, this decreases the S/N by the
square root of the above factors. We create each mock image multiple
times with different noise realizations in order to estimate the
scatter in the fit parameters. Since these mock images are exact
S\'ersic profiles, a single component S\'ersic profile fits
well. However, as images are degraded, the errors in the S\'ersic
index increases, to $\sim5$-$10\%$ at a factor of $10$ degradation.

The exponential and de Vaucouleurs fits to mock 
galaxies with $n\ne 1$ or $n\ne 4$ show interesting systematic
changes in \Reff. Since we have chosen to fix the S\'ersic index of
the bulge, we perform these tests to examine the systematics of
fitting a fixed $n$ profile to an arbitrary S\'ersic profile. If the
S\'ersic index of the model is too low, the 
fitted value for \Reff{} systematically increases if the resolution is
degraded and decreases if the S/N is degraded. The opposite occurs if
the S\'ersic index of the model is too high. The magnitude of these
two effects is $\sim10$-$15\%$, and roughly equal. For a galaxy
`moved' to a higher redshift, the effects should approximately
cancel. These trends are clearly present in the real galaxies of
Sample A as well and are shown in Figure \ref{fig:NReff_SNRES_comp},
which shows the change in \Reff{} as a function of original S\'ersic
index. The cause of these trends is explained by the shape of
S\'ersic profile. For high S\'ersic index, the profile has a sharp
peak at $R=0$ and falls slowly at large radius. Therefore, when
fitting a profile with a too low S\'ersic index to a mock image, the
model underestimates the flux in the center and overestimates the flux
in the outskirts. The opposite is true for a model with a S\'ersic
index that is too large. As the resolution is decreased, the peak at
$R=0$ is smoothed out, and $\chi^2$ is dominated by the flux at large
radius. This will increase(decrease) \Reff{} of a model with a S\'ersic
index smaller(larger) than the actual galaxy. As the S/N is decreased,
the flux at large radius becomes statistically undetectable, and
therefore the residuals in center of the galaxy dominate in the
$\chi^2$, which lead to smaller(larger) \Reff{} in the exponential(de
Vaucouleurs) fits, respectively.
\figsnresNReff
\figsnresSersicN
\figsnresNMag

For the real galaxies in Sample A, we have fit images with the same
decreases in S/N and resolution as those listed above for the mock
galaxy images. These factors are a reasonable choice because they span
the properties of our entire sample. A galaxy with the median model
magnitude in Sample A ($m_r = 13.25$) would have to be moved $\sim 8$ times
farther away to be below the SDSS spectroscopic sample magnitude limit. The ratio between the median size (as measured by the SDSS
value for the radius containing 90\% of the Petrosian flux) of
galaxies in Sample A and the faintest ($m_r < 17.5$) galaxies in the
sample is $\sim 4$, well within the range of our degraded Sample A images. Because we do not have the underlying galaxy model
for the real images, we change the resolution by convolving with a 
Gaussian PSF with a width $\sigma=\sqrt(f^2-1)\sigma_{\mathrm{SDSS}}$,
where $f$ is the factor of the  resolution decrease and
$\sigma_{\mathrm{SDSS}}$ is the FWHM of the Gaussian fit to the SDSS
PSF. After convolving with the additional PSF, we add back the noise
that has been smoothed away, in order to keep the S/N of the image the
same as the original. This procedure ignores the fact that
the smoothed noise is correlated. To lower the S/N, we add sky noise
as if the mean sky background changed by a factor of $f^2$, which
decreases the S/N in the sky-dominated parts of the image by a factor
of $f$. This procedure is no different than that for the mock galaxy
images.  

For the single-component fits, the results of decreasing the
resolution and S/N  can be seen in Figures \ref{fig:NReff_SNRES_comp},
\ref{fig:N_SNRES_comp}, and \ref{fig:NMag_SNRES_comp}. As with the
mock galaxy images, the 
\Reff{} is sensitive to the S\'ersic index of the
galaxy. Figure \ref{fig:NReff_SNRES_comp} shows the fractional change in
\Reff{} as a function of S\'ersic index fitted at high resolution. The different markers indicate different fractional
changes in S/N and resolution. The top row in this figure shows the
change in \Reff{} of a fit with a free S\'ersic index. Decreases in 
resolution(S/N) lead to decreases(increases) in \Reff. Since the
S\'ersic index and \Reff{} are covariant, this trend is also apparent in
Figure \ref{fig:N_SNRES_comp}, which plots the S\'ersic index from the
degraded images against the originally measured S\'ersic
index. Unsurprisingly, the trend in resolution is most pronounced for
high S\'ersic index galaxies, which are the most peaked at
$R=0$. The magnitudes of the models fit at decreased resolution and
S/N are covariant with \Reff, although the change in magnitude is not
large, with the magnitude difference between $-0.06$ and $0.08$, while
the \Reff{} changes by as much as a factor of $2$. The difference in
magnitude between the exponential and de Vaucouleurs model fits to the
same galaxy can be as large as $0.8$. Therefore, the measured
magnitude is reasonably robust to changes in resolution and S/N.

\figsnresBTTMag
\figsnresBTT
The results of changing S/N and resolution for single
component model fits can be understood from the properties of S\'ersic
profiles and the effects lowering the S/N and resolution  have on the
outskirts and centers of the profiles as explained above. However, the
effects of resolution and S/N on B+D fits are more complicated, and the
systematics are less clear. Figure \ref{fig:BTTMag_SNRES_comp} shows
the change in model magnitudes as a function of $B/T$ for the two B+D
models. While the $n_b=4$ B+D model fits show no trend, the scatter is larger
than the scatter in magnitude for the single component fits. The
$n_b=1$ B+D fits show the same trends in the magnitude with $B/T$ that
the single component exponential fits show in magnitude as a function
of S\'ersic index. This is unsurprising, as the $n_b=1$ B+D fits are
a linear combination of 2 exponential fits and should be subject to
some of the same systematics. Figure \ref{fig:BTT_SNRES_comp} shows no
clear trends in $B/T$ with decreased resolution or S/N. The median
fractional errors in $B/T$ for the $n_b=4$ B+D fit as the S/N
decreases by factors of 2, 4, and 10 are $0.01$, $0.03$, and $0.05$. For
a decrease in resolution by the same factors, the errors are $-0.01$,
$-0.02$, and $-0.11$. For the $n_b=1$ B+D fits, the equivalent errors
are $-0.05$, $-0.10$, and $-0.13$ for decreasing S/N, and $0.06$,
$0.13$, and $0.50$ for decreasing resolution. The large change in B/T
for the $n_b=1$ B+D fits can be seen in the lower right panel of Figure 
\ref{fig:BTT_SNRES_comp}. As the resolution decreases, the fitter has
more difficulty distinguishing between the two $n=1$ S\'ersic profiles
in the exponential bulge model. Therefore, small values of $B/T$ will
be scattered upwards as more disk light is confused with the bulge.  
\figSNRESRbulge
\figSNRESRdisk

For the $n_b=1$ B+D, the bulge and disk scale lengths behave the same
way the single exponential scale lengths does; as the resolution(increases)
decreases \Reff{} increases(decreases). The effect is especially
dramatic for the $n_b=1$ bulge; the median error in the bulge \Reff{} for
a change in resolution by a factor of $10$ is $170\%$. This is shown in Figure
\ref{fig:Rebulge_SNRES_comp} for the bulge and Figure
\ref{fig:Redisk_SNRES_comp} for the disk. For the $n_b=4$ B+D fits, the
bulge and disk scale lengths change with the opposite sign for decreasing
resolution, but the additional noise mainly adds scatter. The rms error
in the $n_b=4$ B+D model bulge(disk) \Reff is $16\%$($7\%$), when the
S/N is decreased by a factor of $2$.

Figures \ref{fig:SNRES_exA}, \ref{fig:SNRES_exB}, \ref{fig:SNRES_exC},
and \ref{fig:SNRES_exD} show examples
of $4$ different galaxies and their fit parameters at decreased S/N
and resolution, and increased redshift (see
\S\ref{ssec:redshift}). Although these galaxies follow the general
trends explained above, these trends are not robust. Figures
\ref{fig:SNRES_exC} and \ref{fig:SNRES_exD} show two edge-on disk
galaxies, with similar $B/T$, however, the trends in $B/T$ move in
opposite directions for decreasing S/N and resolution. These figures
demonstrate the difficulty in establishing the systematic behavior of
the model parameters with changes in resolution, S/N, or
redshift. They also show which parameters are the most reliable; the total
magnitudes never vary more than $0.2$ mags, while the $B/T$ changes by
$\pm40\%$. Points in the figures show the median and inter-quartile
range of multiple fits to the same galaxy, with different noise realizations.
\figSNRESexA
\figSNRESexB
\figSNRESexC
\figSNRESexD

\subsection{Redshift}
\label{ssec:redshift}

In addition to varying the S/N and resolution separately for the
galaxies in Sample A, we also modify the galaxy images to create mock
images of the same galaxy at higher redshifts. These redshifted images
involve both a decrease in resolution and S/N as well as a
k-correction. For the redshifted galaxies, we rebin the images to
decrease the resolution, thus keeping the PSF the same size in
pixels. In \S\ref{ssec:SN_RES}, we simply convolved the image with a
larger PSF. We do not k-correct the bulge and the disk separately, but
instead use a single k-correction for the entire galaxy from
\verb+kcorrect+ \citep{Blanton07}. In order to compare the results of
changing redshift to the results of changing S/N and resolution, we
move the galaxies in redshift such that their angular size decreases
by factors of $1.25$, $1.5$, $2$, $4$, $5$, $7$, and $10$. Figure
\ref{fig:NMag_Z_comp} shows the change in magnitude as the images are
moved to higher redshift. This figure is analogous to Figure
\ref{fig:NMag_SNRES_comp} for decreased resolution and S/N. With the
exception of a few galaxies, the errors in magnitude for the
exponential and S\'ersic fits are small. The trend in de Vaucouleurs
fits has the same sign at low $n$ as the trend in magnitude with
decreasing resolution, making the high redshift galaxies slightly too
bright. This is mainly due to an overestimate of flux in the
central regions by a de Vaucouleurs profile. 
\figZNMag
\figZSersicN

Figure \ref{fig:N_Z_comp} shows the trend in measured S\'ersic index
with increasing redshift. The error in S\'ersic index for a factor of
$2$ increase in distance is small. However, the general trend is for
the S\'ersic index to decrease with redshift, which also occurs as the
resolution decreases, suggesting that for SDSS, the measurements of
small galaxies are primarily resolution limited. Figures
\ref{fig:SNRES_exA} through \ref{fig:SNRES_exD} also show the changes
in the measured parameters with increasing redshift (blue lines). As
with the single-component fits, the redshifted images are more similar
to the low resolution images than the low S/N images described in the previous
section. However, the scatter is typically larger for the redshifted
galaxies. For the $n_b=4$ B+D model, the median fractional error in
$B/T$ is $-0.06$, $0.05$, and $0.18$ for decreases in angular size by
factors of $2$, $4$, and $10$. The equivalent errors in the $n_b=1$ B+D
models are $0.12$, $0.35$, and $1.26$ (See Figure
\ref{fig:BTT_Z_comp}). These values are all larger 
than those found by changing the resolution and S/N separately. The
reason for this is two-fold: the redshifted galaxy 
images are rebinned such that the size in pixels becomes smaller,
making the scale lengths more difficult to measure, and the redshifted
galaxies are made dimmer and k-corrected which decreases the S/N more
than simply increasing the sky counts, especially in the brightest
parts of the galaxy. Although the $B/T$ measurements at increasing
redshift are dominated by scatter, the general trend in an increase of
B/T with increasing redshift, in agreement with other studies
\citep{Gadotti08, Simmons08}. 
\figZBTT

\subsection{Limits on fits and systematics}
\label{ssec:limitfits}
The tests on mock and real galaxy images above suggest that limits can
be placed on the resolution and S/N of images it is possible to
fit. One way to quantify the limits on the fits is to examine the
relative errors in the fits parameters (\Reff, S\'ersic index, etc.),
to the relative errors in the flux. For the bright, well-resolved
galaxies in Sample A, the median relative errors of \Reff{} and $n$
for single-component S\'ersic fits are $\sigma_{\Reff}/\Reff /
\sigma_{\mathrm{mag}} \approx 1.5$, and $\sigma_{n}/n / \sigma_{\mathrm{mag}}
\approx 2$. This is in rough agreement
with the ratio of uncertainties from linear calculations
(Gunn, priv.\ comm.). As the resolution decreases, the uncertainty
ratios increase sharply. Thus the determination of \Reff{} is
increasingly uncertain (compared to the flux determination) at low
resolution. The same trend holds for the S\'ersic index measurement
error for a single component S\'ersic fit. In this case, the relative
error of the S\'ersic index increases a factor of $6$ over the flux
error as the resolution decreases by a factor $10$. In order to keep
the relative error in \Reff{} and $n$ comparable to the relative error
in flux (which is set by the S/N), \Reff{} should be $\ga
0.5\times\mathrm{FWHM}$. It is also important to note that the
S\'ersic index, \Reff{}, and the total luminosity are highly
covariant; small errors in any one of these quantities correlates with
errors in the other two. For example, under-estimating the S\'ersic
index due to low S/N will lead to underestimated values for \Reff{}
and the total luminosity. The two-component fits are, of course, more
complicated with even more covariances, and the systematics with S/N
and resolution are not easily discerned, but the limit on \Reff{} of
$0.5$ FHWM is a reasonable choice for bulges and disks in the
2-component fits. In our sample of B+D modeled galaxies, 
$95\%$ of bulges(disks) have $\Reff > 0.4(2.0)\times\mathrm{FWHM}$,
suggesting that most bulges are well enough resolved to be adequately
measured. 

\section{Analysis of Fits}
\label{sec:analysis}
Although we have models for all the galaxies in the sample, for the
further analysis we will remove galaxies for which either B+D model
fails. This occurs in $5\%$ of the sample which leaves $75,357$
galaxies. We also remove all galaxies for which any of single
component fits have a $S/N < 3$ within \Reff, leaving $71,825$
galaxies. All of the removed galaxies are, on average, intrinsically
fainter ($\Delta M_r \approx 0.84$) and 
bluer ($\Delta (g-r) \approx -0.1$) than the remaining sample. This
suggests that these galaxies are often small, irregular star-forming
galaxies for which our models are poor. Each of
these galaxies is fit by five different 
models, and, in order to study bulges, we will need a metric for
choosing which model to use for each galaxy. Although our
choice is informed by the the statistical quality of the fit, we
also introduce physical interpretations of the data in order to help
categorize our sample. Therefore, although for many galaxies the
choice of model is obvious, the methods for choosing models are not
definitive and can be changed to suit the problem at hand. Our main
interest is in identifying galaxies with classical ($n_b=4$) bulges,
which is reflected in our method of categorizing galaxies described below.

\subsection{$n_b=1$ vs. $n_b=4$ Bulges}
\label{ssec:n1n4bulge}
The obvious metric for choosing the best-fit model for each galaxy is
the $\chi^2$ value from the fitter. However, the $\chi^2$
values for different models do not differ greatly, and many of the
galaxies have significant structure (spiral arms, rings, etc.) that is
poorly fit by the models, leading to large values for $\chi^2$. Thus, we
require 
additional information. Instead of relying on statistical tests alone, we
categorize galaxies by using a series of tests similar to the logical
filter employed by \citet{Allen06}. As in their work, our default
choice will be a single component S\'ersic model. We first separate
out the galaxies which statistically are best fit by single component
exponential or de Vaucouleurs profile. In addition, we
compare the disk and bulge components of the B+D fits to the
exponential and de Vaucouleurs fits. If the 2-component fits and
1-component fits are the same to a chosen tolerance (i.e. either the
bulge or disk flux is consistent with zero), we model the galaxy with a single
component. Finally, we divide the galaxies into $n_b=4$ and $n_b=1$
bulges, using the colours and axis ratios of the bulges.

\begin{enumerate}
\item \em{Exponential disk profiles: }\em  The easiest category 
  to find is the
  pure disk galaxies. Because exponential disks are used in both B+D
  models, these models and the pure exponential model yield similar
  results for pure disks. The first 
  criterion we use to select 
  pure disk galaxies is the statistical significance of the bulge
  component. We do this by comparing the (non-reduced) $\chi^2$ value
  of the B+D fits to the $\chi^2$ from single component exponential
  disk fits. Since the 
  B+D fits contain more parameters, the quality of the fit should
  improve and the $\chi^2$ of the fit should decrease. For linear
  models fit to data with Gaussian errors, the expected 
  difference in $\chi^2$ can be quantified \citep{Lupton93}. The B+D
  model fits an image with $n$ pixels using $10$ free parameters. An
  exponential disk uses the same model, but fixes the bulge
  parameters, thus increasing the numbers of degrees of freedom by
  $4$. If these were linear models, the difference of the $\chi^2$ for
  the models, $\chi^2_{\mathrm{exp}} - \chi^2_{\mathrm{B+D}}$ would be
  independent and follow a $\chi^2$ distribution for $4$ degrees of 
  freedom. The relevant statistic is
  \begin{equation}
    \label{eq:ftest}
    f=\frac{\left(\chi^2_{\mathrm{exp}}-\chi^2_{\mathrm{B+D}}\right)/4}{\chi^2_{\mathrm{B+D}}/(n-10)}\quad ,
  \end{equation}
  where $f$ follows an $F_{4,n-10}$ distribution
  \citep[see][]{Lupton93}. If $f$ is large, 
  there is a low probability of the difference in $\chi^2$ occurring
  due to the additional degrees of freedom alone. Although our models
  are not linear and our errors are only approximately Gaussian, we
  require that $f$ be 
  larger than the $99.99\%$ level from the $F-$distribution. This
  ensures that the bulge is a statistically significant part of the fit.
  We compare the exponential fit to each B+D fit separately, and
  assign galaxies for which neither bulge is statistically significant
  to the exponential disc category. There are $5,893$ galaxies which
  satisfy this criterion.
  However, we would also like to include galaxies for which the bulge
  may be formally statistically significant, but for which the bulge
  is small compared to the systematic errors examined in the sections
  above. In order to do this, we select galaxies for which the disk in
  the B+D model matches the single exponential fit in total
  flux, axis ratio ($q_d$), and \Reff{} to within $10\%$. When these galaxies are included
  in the sample, there are $15438$ pure disk galaxies. Only $2.6\%$
  of the $5893$ galaxies which pass the first (statistical) test, fail
  the second test.
\figdVcQ
\item \em{de Vaucouleurs profiles: }\em The initial procedure to select de
  Vaucouleurs, or elliptical galaxies is similar to that used for
  exponential galaxies. However, in this case we can only directly
  compare the single de Vaucouleurs profile fit to the $n_b=4$ B+D
  model. First, we select $755$ galaxies for which the difference in $\chi^2$
  values for the de Vaucouleurs model and the $n_b=4$ B+D model is not
  statistically significant. The $4$ galaxies which satisfy
  both the conditions for a de Vaucouleurs profile and an exponential
  profile are removed from both lists, and fit by a single
  S\'ersic component instead.

  We expect the number of elliptical
  galaxies in our sample to be much larger than $755$. The reason this
  number is so low can be seen in Figure
  \ref{fig:example_profs}d. Many elliptical galaxies are well fit 
  by an $n_b=4$ bulge and a large, low surface brightness disk. Because
  the disk has a large scale length, the measured $B/T$ can be as low
  as $\sim40-50\%$. This has been found in other studies
  \citep{Allen06}, and is typically not interpreted as a low
  luminosity disk, but rather as an extended halo,  a
  second S\'ersic component \citep{Gonzalez05}, tidal features, or a
  S\'ersic index that is larger than $4$. This feature could also be
  due to uncertainties in the sky in the vicinity of bright, extended
  objects, although the new DR8 reductions attempt to reduce these
  uncertainties. From the 
  $755$ galaxies above, we find that $90\%$ of them have $g-r>0.55$
  and $g-i>0.8$, and $b/a > 0.55$. In order to find elliptical
  galaxies, we select galaxies which
  satisfy these three criteria using the de Vaucouleurs models. These
  galaxies all lie on the red sequence, but not all galaxies on the
  red sequence are ellipticals \citep{Bundy2010}, and S0 galaxies
  should be modeled as B+D galaxies. Therefore, we must
  also examine the two component fits for these red sequence galaxies. Typically,
  ellipticals in our sample are well fit by $2$ round $n=1$ S\'ersic
  components, or by a large $n_b=4$ bulge and a very large, low surface brightness
  exponential component. If the galaxies along the red sequence have
  an $n_b=1$ B+D fit with a round ($q_d > 0.4$), red ($g-r > 0.65$)
  disk, and have an $n_b=4$ B+D fit with a large ($B/T > 0.4$), red
  ($g-r > 0.65$) , round ($q_b > 0.55$) bulge, we assume they are
  elliptical galaxies and fit them with single component de
  Vaucouleurs model. This category contains $9692$ galaxies. Figure
  \ref{fig:dvc_q} shows the axis ratio distribution for the $n_b=4$
  B+D models for the de Vaucouleurs galaxies. If these galaxies were
  dominated by S0 galaxies with real disks, the $q_d$ distribution
  would be flat. Instead, the distribution is biased to high values of
  $q_d$ with a peak at $q_d \approx 0.75$. 

\item \em{bulge + disk: }\em After removing all galaxies which are
  best-fit by a single exponential or de Vaucouleurs profile, there
  are $46,155$ galaxies remaining. Among these galaxies, we select
  only those which have physically sensible bulge and disk parameters
  for both models. We require that the bulges and disks in both models
  are detected in $g$,$r$,and $i$-band images. This eliminates 6366
  galaxies. We also require that the bulges and disks are arranged in
  the expected manner, i.e. the bulge \Reff{} is smaller than
  that of the disk and the bulge flux dominates in the central part of
  the galaxy. This leaves $24,974$ galaxies.

  At this point, we must distinguish between galaxies with classical,
  $n_b=4$ bulges and pseudo-bulges ($n_b=1$). For most galaxies, the
  $\chi^2$ values for the fits do not differ significantly. From the
  $\chi^2$ values for the fits, we can calculate the probability that
  a model is a good fit using equation \ref{eq:chisqCDF}. For Gaussian
  errors, the expected probability value for a good fit is $50\%$; a
  value near $0$ represents a poor fit, while a value near $100\%$
  suggests the model over-fits the data, or the errors are
  underestimated. Only $6\%$ of the galaxies have B+D model fits with
  probability values between $5$ and $95\%$, suggesting that the
  $\chi^2$ value is not useful for discriminating between B+D
  models. 

  The distributions of bulge colours and inclination angles for
  all B+D galaxies fit with both models are shown in Figure
  \ref{fig:bulgeGR}. The colour 
  distributions for bulges peak around 
  $g-r\sim0.8$, which lies along the red sequence, but there is an
  excess of bulges blue-wards of $0.8$. The existence of this excess
  was also found by \citet{Allen06}. The expectation is that
  pseudo-bulges should be disk-like ($n=1$ profiles) \citep{Fisher08}
  and have ongoing star-formation \citep{Fisher06,
    Kormendy04}. Therefore, instead of relying on  the 
  statistical properties of the fits, we categorize bulges
  by their shape and colour. We expect classical bulges to be red and
  round, compared to their disks. If the $n_b=4$ bulge component has
  $g-r > 0.6$ and $q_b/q_d > 0.65$, we model the galaxy with an $n_b=4$
  bulge. Otherwise, if the $n_b=1$ bulge and disk have similar axis
  ratios ($0.5<q_b/q_d<2.0$), we categorize the galaxy as a
  pseudo-bulge ($n_b=1$) host. This leaves a sample of $14042$ $n_b=4$
  B+D galaxies and $6684$ $n_b=1$ B+D galaxies. The remaining $4,275$
  galaxies are placed in the S\'ersic profile category. The exact
  choice for the axis ratio and colour cutoffs for this sample is
  somewhat arbitrary, and changing the values will change the
  categories for some galaxies. The division above is somewhat
  conservative as it places a significant fraction of the galaxies
  into the S\'ersic category.

  Our models do not include a separate bar component and we expect
  light from the bar to typically be included in the bulge. Barred
  galaxies are distinguished by bulges with small axis ratios compared
  to their host disk. Therefore, barred galaxies are categorized either as
  $n_b=1$ B+D galaxies (for weak bars) or S\'ersic galaxies. The
  $3963$ galaxies with bulges significantly flatter than their disks
  (i.e., bars) are placed in the S\'ersic profile category. Of these
  barred galaxies, around $1/2$ have red bulges/bars. 
  \figBulgeGR

\item \em{S\'ersic profile: }\em We use the S\'ersic profile for
  galaxies which do not fit in any of the categories above. This 
  group includes $25,970$ galaxies, and is the largest category. Of the
  galaxies in this group, $40\%$ do not have $\chi^2$ values which are
  significantly improved by adding a second component to the S\'ersic
  fit. The distribution of S\'ersic indices and magnitudes for this
  group are shown in Figures \ref{fig:Cat_N} and
  \ref{fig:Cat_Mag}. The median S\'ersic index for these galaxies is
  $n=1.5$, suggesting these galaxies are disk-like rather than
  ellipticals. Furthermore, the galaxies fit with S\'ersic profiles are
  intrinsically fainter than the majority of our sample; the median
  $r$-band absolute magnitudes differ by $0.5$ magnitudes. These
  galaxies also have a median $g-r$ colour $0.1$ mag bluer than the
  entire sample. All of this suggests that a large fraction of the
  S\'ersic profile fit galaxies are faint, star-forming, disk-like (or
  irregular) galaxies for which a B+D model is a poor choice.
\end{enumerate}

We have evaluated our category assignments by visual inspection on
Sample B, a subset of $265$ randomly selected galaxies. Of the $51$
galaxies classified as exponential disks, $\la 6\%$ are too irregular
to obviously identify a disk. Of the $50$ de Vaucouleurs galaxies, $7$
should be classified as B+D galaxies. There are $19$ $n_b=1$ B+D and
$59$ $n_b=4$ B+D galaxies in this sample. The $n_b=4$ and $n_b=1$
galaxies are difficult to distinguish by eye, but only $\sim 8\%$ ($6$
galaxies) of the sample can be visually classified as ellipticals.
The remaining $85$ galaxies are fit with a S\'ersic profile. As
expected from the S\'ersic index distribution in Figure
\ref{fig:Cat_N}, most of these galaxies are disk galaxies, and many
nearly edge on, or dominated by a large bar. It makes sense that these
types of galaxies are most likely to have a poor quality or unphysical
fit for one of the B+D models, and therefore be placed in the S\'ersic
category. Additionally, the S\'ersic-fitted galaxies are fainter
than the  B+D galaxies, with a median SDSS model $r$-band magnitude of $16.7$
compared to $15.9$ for the $n_b=4$ B+D galaxies.  
\figCatN
\figCatMag

In addition to visual inspection for a small sample, we have also
examined the statistical properties of the whole sample. Figure
\ref{fig:Cat_Mag} shows the fraction of galaxies in each category as a
function of apparent and absolute magnitudes. Unsurprisingly, de
Vaucouleurs galaxies (ellipticals) dominate the bright end of the 
sample. At the faint end, $50\%$ of the galaxies fainter than $\sim 17$ in
$r$ are placed in the S\'ersic fit category, suggesting that B+D
fitting becomes difficult for faint galaxies. The absolute
magnitude distribution shows that the intrinsically faintest galaxies
in our sample are typically categorized either as exponential disks or
S\'ersic fit galaxies. This is in agreement with observations that intrinsically
faint galaxies are more likely to be disorganized, and therefore more
difficult to fit with a B+D model than brighter galaxies. The 
rightmost panel in Figure \ref{fig:Cat_Mag} shows the fractional distributions of
absolute magnitude for the $43,403$ galaxies brighter than $16.7$ in
$r$. This distribution is not significantly different than that for
the total sample; only the fraction of exponential galaxies
decreases. This demonstrates that the high fraction of S\'ersic modeled
galaxies at low luminosity is due in large part to the intrinsic
nature of these galaxies, and not just the inability of the fitter to
fit B+D models to apparently faint galaxies. Thus, our fitter and
categorization algorithm successfully identify B+D galaxies in our
sample. The galaxies which default to a S\'ersic profile are most
likely disorganized, intrinsically faint galaxies for which a B+D
model is inaccurate. Finally, we note that the shape of the absolute magnitude
distribution for the intrinsically bright sample (lower right panel)
is different from the absolute magnitude distribution for the whole
sample, which should not be the case for a truly magnitude-limited
sample. However, our sample also includes a volume (redshift) limit
which affects the distribution of bright galaxies, and we do not
expect the distribution of absolute magnitudes to scale simply.

Figure
\ref{fig:Cat_N} shows the distribution of S\'ersic indices for a
single component fit for the different categories. As expected,
exponential disk galaxies have $n\approx1$, while de Vaucouleurs
galaxies have $n\ga 2$ (by selection), and a wide distribution of
S\'ersic indices peaking at $\sim 4.5$. The B+D modeled galaxies both
have peaks in their S\'ersic index distributions near $n\approx 2$,
but the $n_b=4$ B+D galaxies tend to have slightly higher S\'ersic
indices for the whole galaxy than the $n_b=1$ B+D modeled galaxies. 

\subsection{Comparison to Other Studies}
\label{ssec:compareGadotti}
\citet{Gadotti09}(G09) have performed bulge-bar-disk decompositions on a
sample of $\sim1000$ SDSS galaxies, selected to give high quality
model fits. We have matched their sample to ours and found $442$
matches. Eighty percent of the remaining galaxies in their sample are
outside our redshift range. K-correcting our magnitudes to $z=0.1$, we
find excellent agreement between the model magnitudes, with the
mean $\Delta M_r = -0.11 \pm 0.25$. In galaxies for which we both
measure disks, the scale lengths 
also agree well, although ours are systematically larger, with a median
difference of $8\%$. In comparing the B/T, we find that our
measured B/T is systematically larger by $22\%$. This is not
surprising, since G09 include a bar component in their models and we do
not, which will substantially affect the bulge \citep{Gadotti08}. If
we compare our B/T to their B$+$Bar/T, we find no offset, although
the inter-quartile range of (B/T)$-$(B$+$Bar/T) is $[-34\%,
58\%]$. This helps confirm that our bulges include bars. Similarly,
our bulge effective radii differ significantly. Because we
set the S\'ersic index of the bulge to either $4$ or $1$, our
classical bulges typically have larger scale lengths and our
pseudo-bulges typically have smaller scale lengths than those in
G09. In comparing our classification to G09 (see Table
\ref{table:G09comp}, we find that our use of 2 component fits is more
conservative. Most of the galaxies we classify as classical bulges G09
does as well. However, almost half of the galaxies G09 classifies as
classical bulges, we classify as ellipticals. Similar trends, although
less robust, hold for the classification of pseudo-bulges and
pure disk galaxies. The fraction of S\'ersic galaxies we find in the G09
sample is lower than that in our complete sample. This is due to the
fact that G09 have more carefully selected their sample for galaxies
which can support a B+D model. 
\tabGcomp

We have also compared our sample to the larger sample from
S11. There are $67,333$ galaxies which are fit in both
samples; the missing galaxies are the brightest in our sample which
the S11 sample excludes. The S11 work fits each galaxy
with three models and uses an $F-$test to determine whether a
bulge$+$disk model is needed. A galaxy is fit with a bulge$+$disk if there
is less than $32\%$ chance that a single S\'ersic profile is a better
fit. This threshold is much higher than the one we set for the
$F$-test in \S\ref{ssec:n1n4bulge}, which leads to more galaxies fit
with a bulge$+$disk model in the S11 sample. Nonetheless, the results
of their $F-$test compare favorably to our division into galaxies with
bulges and disks. Seventy-five percent  of the galaxies we fit with a
$n_b=4$ bulge$+$disk also pass their $F-$test, while only $23\%$ of
our S\'ersic-modeled galaxies are better fit with a bulge$+$disk model
according to S11. However, we find poor agreement between the S11
$F-$test and our categorization of elliptical galaxies; fifty percent
of the galaxies we fit with a single de Vaucouleurs profile have a
disk component according to S11. This is probably due to our
inclusion of galaxies with large, low luminosity disks in the
elliptical galaxy category. Finally, S11 find that the SDSS
images are typically not of high enough resolution to determine the S\'ersic
index of the bulge, in agreement with our conclusions
above.

We also compare the measured parameters for galaxies from S11 to our
sample. We limit the comparison to the $\sim 10,000$ galaxies that
both studies assign a $n_b=4$ bulge$+$disk model, and find 
reasonable agreement for all the model quantities. The $r$-band
magnitudes found by S11 are systematically $0.06$ magnitudes brighter,
which is due to differences in the cutoff radii and the sky
subtraction. There is no offset in disk scale lengths between the
studies, and the scatter ($\mathrm{RMS}[(R_{\mathrm{this\ work}}-R_{S11})/R_{\mathrm{this\ work}}]$) is
$11\%$. The differences between the bulge scale lengths are
larger. The bulge scale lengths we measure are $17\%$ larger on
average than those in S11, and
the scatter  is $50\%$. However, the $B/T$
ratios we find are only 
$2\%$ larger on average, with a scatter ($\mathrm{RMS}[B/T_{\mathrm{this\ work}} - B/T_{S11}]$) of $0.1$.  The
fluxes are more robust than the scale lengths and the differences in
bulge and disk colors are similar to the differences in $B/T$.  

\subsubsection{Kormendy Relation}
\label{sssec:kormendy}
Unlike classical bulges, pseudo-bulges are not expected to lie on the
fundamental plane. In fact, the difference in the Kormendy relation
\citep{Kormendy77} for classical and pseudo-bulges can be used to
identify pseudo-bulges (G09). Figure \ref{fig:kormendy} shows the
Kormendy relation for ellipticals and bulges in our sample. The solid
line shows the relation fit to the ellipticals (red contours), which
has the equation: $\langle \mu_{\Reff} \rangle = (2.27\pm 0.04)\log \Reff  + (12.18\pm
0.15)$, which agrees with the Kormendy relation for
SDSS galaxies \citep{Bernardi03}. The central contours for the
classical ($n_b=4$) bulges lie along the same relation as the
elliptical galaxies (at smaller radii), while the $n_b=1$ B+D form a
steeper relation, and the majority do not lie on the elliptical galaxy
Kormendy relation. This supports or use of colour and shape as a
discriminator between classical and pseudo-bulges. 
\figKormendy

\subsection{Inclination and internal extinction}
\label{ssec:inclination}
Just as we must correct for extinction from our own galaxy, the colours
and magnitudes in our sample should be corrected for internal
extinction from the sample galaxies themselves. Since elliptical
galaxies contain little dust, they are not affected by intrinsic
extinction, while disk galaxies become redder and fainter with
increasing inclination \citep{Burstein91}. Additionally, 
even if bulges contribute nothing to the extinction themselves, their
light will be attenuated by the 
encompassing disk.

Before correcting for inclination, we examine the distribution
of inclination angles of disks in our sample. These are shown in
Figure 
\ref{fig:diskIncDist}. The dotted line is the theoretical distribution
of axis ratios, assuming a disk flattening of $1/q_z = 5$. The
measured axis ratio and its probability density function are 
\begin{eqnarray}
  \label{eq:q_pdf}
  &q_d& = \sqrt{q_z^2 + (1-q_z^2)\cos\theta}\ ,\ \mathrm{and} \\ \nonumber
  &P(q_d)& = q/\left[\sqrt{(1-q_z^2)(q^2-q_z)^2}\right]\ ,
\end{eqnarray}
where $\theta$ is the inclination angle. Figure \ref{fig:diskIncDist}
shows that our sample is missing galaxies face-on ($q_d=1$) galaxies,
which is not surprising since any intrinsic asymmetry in the galaxy will
decrease the measured axis ratio. Additionally, the lower panel shows
that the distribution of disk angles depends on the profile used for
the bulge. The thick lines show the distributions of disk axis ratios
for the {\it same} set of galaxies using an $n_b=4$ and an $n_b=1$ B+D
model. The measured disk axis ratio tends to be higher (less inclined)
for the $n_b=1$ B+D models. We have examined a set of galaxies for
which the difference in measured inclination angle for the two B+D
models is large. Many of these galaxies are edge-on disks with
prominent bulges. These galaxies are poorly fit by the $n_b=1$ B+D
models, as the disk and bulge have the same profile shape and are
easily confused. Nonetheless, because our $n_b=1$ versus $n_b=4$ bulge
separation relies on colour, and highly inclined disks will lead to
redder colours, we will re-categorize the B+D galaxies, using
inclination-corrected bulge colours.

The internal extinction correction can be divided
into two parts, the increase in extinction due to inclination, and the
extinction due to dust in face-on galaxies. The former portion can be
corrected by removing trends in galaxy properties with inclination
\citep{Tully98,Driver07,Shao07,Maller09}. \citet{Driver07} and
\citet{Driver08} have derived extinction corrections for bulges and
disks separately using B+D decompositions of the Millennium Galaxy
Catalogue \citep{Liske2003}. Their magnitude corrections also include a face-on
attenuation. \figdiskIncDist
In this work, we will only address the inclination dependent portion
of the internal extinction. In order to compare the properties of
classical bulges in this sample to elliptical galaxies, we will be
required to address face-on attenuation, but in comparing the
properties of bulges to other bulges, it is not necessary. In order to
correct the colours for inclination effects, we use the corrections
derived by \citet{Maller09}(M09). These corrections are derived by
removing trends in $\lambda-K$ colours with axis ratio ($q$). The corrections
to the $\lambda-K$ colours are of the form
\begin{eqnarray}
\label{eq:mallerCorrect}
&A_{\lambda-K} &= -\gamma_\lambda\log q\quad \mathrm{where} \\ \nonumber
&\gamma_\lambda&=\alpha_o + \alpha_K(M_K+20)+\alpha_nn_{\mathrm{Sersic}}\ ,
\end{eqnarray}
where $M_K$ is the extrapolated $K_s-$band magnitude from 2MASS
\citep{Jarrett00}. The fitting constants $\alpha_x$ can be found in
Table 2 of M09. Just over $50\%$ of the galaxies in our sample can
be matched to a 2MASS object. For the remainder of the objects, we can
estimate $M_K$ from the SDSS model $i$-band magnitude and $r-z$ colour as:
\begin{equation}
\label{eq:kbandfit}
M_K = 1.66 + 1.07\times M_i - 0.94\times(r-z)\quad ,
\end{equation}
which is derived by a least-squares fit to an apparently bright ($m_r
> 15.7$) sample of galaxies with 2MASS data.  The rms scatter around this relation is less than $0.002$ magnitudes. Below, we examine
the correction applied to galaxies with and without $K$-band
observations separately. We have also tested our inclination
correction on a bright ($m_r > 15.7$) sample of $\sim 23,000$ galaxies,
of which $95\%$ have $K$-band observations. These galaxies are on
average redder than the rest of our sample, but the results of the
inclination correction remain the same. 

The inclination
corrections in M09 were derived using a sample of 
$\sim8000$ disk-dominated galaxies from SDSS, most of which are
included in our sample. These corrections were derived for galaxies
fit with a single S\'ersic profile. For our sample of B+D fits, we
apply a correction both to the  bulge and the disk, separately. We use the
single component S\'ersic fit for $n_{\mathrm{Sersic}}$, the total
$K-$band magnitude (or an estimate), and the measured disk inclination
angle for the correction. Because we use the disk inclination instead
of the total galaxy inclination, we find that the M09 $A_{\lambda-K}$
over-corrects the disk colours in our sample, both for bulge-less
galaxies and galaxies with bulges. We have adjusted the corrections
for this by removing trends in disk colour with
inclination. Additionally, we have assumed that only $1/2$ the bulge
light is extincted by the disk. Our justification for this is the
assumption that the bulge light extends well above the disk, gas, and
dust. Therefore, if all the extinction occurs in the the plane of the
disk, a reasonable assumption, a significant fraction of the bulge
light will not be extincted, and the bulge correction should reflect
this. The form the the fractional bulge correction is
\begin{equation}
\label{eq:bulgeAlambdaK}
A_{\lambda-K}(\mathrm{bulge}) = 2.5
\log\left(0.5\left(1+10^{0.4A_{\lambda-K}(\mathrm{disk})}\right)\right)\ .
\end{equation}
\figMallerExp
The M09 inclination correction applied to our sample of pure disk
galaxies is shown in the first two panels of Figure
\ref{fig:maller_exp}. These panels show that the M09 corrections
over-correct inclined disks. They also show that the distribution of $g-r$
colours for galaxies without $K$-band observations is bluer than
that for galaxies with $K$-band observations (about $1/3$ of disk
galaxies). This makes the difference between inclined an face-on
galaxy colours small. Indeed, the difference in median colour between between
inclined galaxies and face-on galaxies without $K$ is less than $0.1$
mags.

The over-correction of disk colours using M09 also occurs
for disks in B+D modeled galaxies. This over-correction is partially
due to the fact that we use the disk inclination instead of the total
galaxy inclination in the equation \ref{eq:mallerCorrect}. We correct
for this by removing trends in disk colours with $\log q_d$,
adjusting the $\alpha_0$ term in equation \ref{eq:mallerCorrect}. We use our
sample of pure disk galaxies, including galaxies with and without
$K$-band observations. We have
chosen to correct colours including $i$-band because the M09
correction works best for the $i$-band (among $g$, $r$, and
$i$). This is evident from the trend in $i-K$ with the $K$-band
magnitude for disk galaxies. The M09
correction aims to remove trends in this relation with $\log q_d$;
examining our data, the trend with the slope of the relation between
$i-K$ and $K$ is smaller than that for $g$ and $r$. Given this choice,
the $i$-band $\alpha_0$ does not change. For the $u$, $g$, $r$, and
$z$ bands, the values $-0.302$, $-0.183$, $-0.089$, and $0.050$ are
added to $\alpha_0$. The results of this change to $\alpha_0$ are shown in the
right two panels of Figure \ref{fig:maller_exp}, for galaxies with
large ($q_d < 0.4$) and intermediate ($0.4 < q_d \leq 0.55$)
inclinations. The medians of the corrected colour distributions for
inclined galaxies are still $\sim 0.01$ mag too blue, but this should
be compared to the $\sim 0.05$ mag offset in colour for the
uncorrected M09 inclination corrections. 

Figures \ref{fig:maller_BDa} and \ref{fig:maller_BDb} shows the
inclination correction applied to B+D fitted galaxies. Here, the
corrections include our modifications, and the bulge correction is
given by equation \ref{eq:bulgeAlambdaK}, which only extincts $1/2$
the bulge light. For the $n_b=4$ B+D galaxies, the bulge and disk
corrections work reasonably well for galaxies with large and
intermediate inclinations. If instead of correcting $1/2$ the bulge
light, we correct all of it, the median corrected $g-r$ of the
most inclined galaxies is $0.02$ magnitudes bluer than the median colour
of bulges in face-on galaxies, while they agree to $0.002$ magnitudes
when only corrected $1/2$ the bulge light. The distribution of bulge
and disk colours for $n_b=1$ galaxies are not as successful. For these
galaxies, the bulge colours are under-corrected, on average, while the
disk colours are over-corrected. This is especially true for galaxies
without $K$-band data. Like the exponential disks without $K$-band
observations, these galaxies have blue disks ($g-r \approx 0.4$) no
matter what  the inclination angle, and are probably not greatly
affected by dust or extinction. For the $n_b=1$ bulges, the colour distribution
is wide, especially for inclined disks (see Figure \ref{fig:maller_BDa}, upper right
panel). Some of this spread is due to uncertainties in the bulge
colours for small bulges.  The $n_b=1$ B+D galaxies are
intrinsically fainter than the $n_b=4$ B+D galaxies (median
$M_r$ of $-19.4$ versus $-20.4$), which further implies they will be
less affected by dust and the relation between colour and inclination
will be weaker. 
\figMallerBDa
\figMallerBDb

Using the colour corrections from this section, we can redo the
categorization from \S\ref{ssec:n1n4bulge}. This does not affect the
results greatly. In the whole sample, about $200$ galaxies are
recategorized. For the most part, galaxies are removed from the de Vaucouleurs
category and placed in the B+D and S\'ersic categories. The $n_b=4$,
$n_b=1$, and S\'ersic categories grow by $39$, $64$ and $48$ galaxies,
respectively. 

\subsection{AGNs}
\label{ssec:AGN}
\figAGNCompare
Because we have chosen to fit the entire low redshift, spectroscopic \verb+VAGC+
catalogue, our sample contains nearby active galactic nuclei
(AGNs). If an AGN is sufficiently bright, it may contaminate the B+D
fit, especially the colours of the bulge. However, most nearby AGN do
not outshine their host galaxy, and the B+D fits and colours will not
be greatly affected \citep{Hao05b,Simmons08}. \citet{Simmons08} find that
the $B/T$ for AGN hosts is systematically over-estimated by
$\sim10\%$, but for weak AGN ($L_{AGN}/L_{\mathrm{host}} \la 10\%$),
the change in $B/T$ is consistent with zero. \citet{Gadotti09} find
that including a point source (AGN) component in their fits is not statistically
significant. \citet{Kauffmann03} find that, although host galaxies
of powerful AGN have undergone more recent star formation, the
structural properties of narrow-line AGN hosts and the general
early-type galaxy population are the same.

Broad-line AGN, however are
associated with a bright, blue continuum source which may contaminate
our sample. In order to check on contamination from AGN, we have
matched our sample against the (updated) broad-line AGN (BLAGN) sample from
\citet{Hao05a}(Strauss, priv.\ comm.).  This is a sample of $10,015$ BLAGN
spectroscopically selected from SDSS DR7. Matching these samples gives
$1134$ galaxies with AGN in our sample, or $1.3\%$. This is comparable
the fraction of galaxies with broad-line AGN \citet{Hao05a} found in
their much larger spectroscopic sample. Of these $1134$ galaxies,
$205$ are classified as $n_b=4$ B+D galaxies and $133$ are classified
as $n_b=1$ B+D galaxies. The remaining sample is divided into $54$ de
Vaucouleurs galaxies, $289$ exponential galaxies, and $453$ S\'ersic
galaxies. Figure \ref{fig:agn_compare} compares the
properties of the BLAGN fitted with an $n_b=4$ B+D to the sample of
galaxies identified as B+D galaxies in \S\ref{ssec:n1n4bulge} (also
all fitted with $n_b=4$ B+D models). The median bulge
\Reff{} for the BLAGN is $2.2\times\mathrm{PSF\ FWHM}$, while for the
whole sample it is $1.9\times\mathrm{PSF\ FWHM}$. However, both of
these values are larger than the HWHM of the PSF, suggesting for the
BLAGN sample, the bulge component is not purely based on the AGN. If
anything, AGN tend to be hosted in large bulges. Using the $n_b=1$
models, the median $g-r$ colours of the BLAGN bulges are $0.05$ mag bluer than the 
underlying sample, suggesting some contamination from the
AGN. Nonetheless, BLAGN make up a very small fraction of our sample,
so our results are not greatly affected by this contamination. BLAGNs
are not even the majority of extremely blue bulges, which are dominated
by galaxies with dense, central starbursts.

\subsection{Proxies for B/T and morphology}
\label{ssec:proxies}
Although we have performed B+D decompositions for a large number of
nearby galaxies, these fitting methods are time-consuming, and
inappropriate for higher redshift, lower resolution data. Therefore,
we compare our results to other  methods for determining
galaxy type and $B/T$. Two simple measures of galaxy type include the
global S\'ersic index and the Petrosian concentration, $C$. As with
most studies using SDSS, the concentration is taken as
$r_{90}/r_{50}$, where $r_{xx}$ is the radius of the circular
containing $90(50)\%$ of the Petrosian flux. Unlike the $B/T$ or the
S\'ersic index, concentration measurements do not take seeing into
account. However, $99\%$ of the galaxies in our sample have $r_{50}
> \mathrm{PSF\ FWHM}$, so the concentration is not
significantly affected by seeing in our sample. For perfect S\'ersic
profiles, concentration is a function of axis ratio, $q$, and S\'ersic
index, $n$; for $q=1$ and $n=4(1)$, $C\approx3.2(2.4)$ \citep{Blanton03}.
\figBTTConc
\figBTTN

The top panel in Figure \ref{fig:BTT_Conc} shows the relation of Petrosian
concentration to $B/T$ for the B+D fitted galaxies. The Spearman's
rank correlation coefficient is $0.73$. Fitting a line 
to the inverse relation (for $B/T < 0.6$), yields
\begin{eqnarray}
\label{eq:concBT}
&C&=(1.45\pm0.01) B/T + 2.21\pm0.004 \ ,\\
&C&=(1.41\pm0.02) B/T + 2.22\pm0.005 \ ,\mathrm{\ for\ n_b=4,}\nonumber \\
&C&=(1.34\pm0.02) B/T + 2.22\pm0.005 \ ,\mathrm{\ for\ n_b=1. }\nonumber
\end{eqnarray}
None of these relations are as steep as those found by
\citet{Gadotti09}, who finds a slope of $1.93$. The difference in
slope is due to the restriction of their sample to galaxies with $b/a > 0.9$, which
typically have lower measured $B/T$ values. If we restrict our sample
to galaxies with $b/a > 0.9$, the measured slope increases to
$2.04\pm0.07$. The lower panel in
Figure \ref{fig:BTT_Conc} shows the distribution of concentrations for
the different galaxy types. Galaxies fit by either an exponential or a
de Vaucouleurs profile have well-separated concentrations, centered
around $2.2$ and $3.1$, respectively. B+D galaxies have concentrations
in between these values. Figure \ref{fig:BTT_N} shows the relation
between S\'ersic index and $B/T$; the correlation coefficient is
$0.52$. The S\'ersic index used is that from 
our single component S\'ersic fit, although using the S\'ersic index
from \citet{Blanton03} does not significantly change the
results. There is a significant scatter to high S\'ersic index for
B/T$\approx 0.3$. This is possibly analogous to the two branches of
$B/T$ seen for high S\'ersic index galaxies in
S11. They suggest the scatter is due to poor separation
between double and single component galaxies. High S\'ersic index
models are probably not good fits to the underlying galaxies,
highlighting the difficulty in  
using S\'ersic index as a B/T indicator. Like
\citet{Gadotti09}, we find that concentration is a better predictor of
$B/T$ than S\'ersic index.

\subsection{Properties of blue, green, and red galaxies}
\label{ssec:bgr_gals}

The bimodality of galaxy colours has been well-established both at low
redshift \citep{Strateva01,Blanton03,Baldry04} and to $z\approx 2$
\citep{Bell04}. In optical colours, the bimodal distribution
can be modelled as two Gaussian distributions
\citep{Strateva01,Baldry04,Mendez11}. The minimum between these two
distributions, the `green valley,' is thought to contain galaxies in
transition between the red sequence and the blue cloud. The
transition occurs as star formation is quenched and different
mechanisms for star formation quenching (harassment, ram pressure
stripping, AGNs, merging) can have profound effects on
morphology. \citet{Mendez11} have found that the morphological
properties of green valley galaxies at $0.4\le z\le 1$ are
intermediate to those on the red sequence and blue cloud, and do not
show an enhancement of merger activity, indicating star formation may
be quenched by internal processes and less dramatic environmental
factors.  
\figCMDoverlay
\figCMDoverlayBD

\citet{Strateva01, Driver06} show the galaxy colour
bi-modality can be traced to the colour bimodality of
bulges(ellipticals) and disks, not separate types of galaxies. Figures
\ref{fig:CMD_overlay} and \ref{fig:CMD_overlayB} show the colour
magnitude diagrams for our sample, both for whole galaxies, and
galaxies divided into bulges and disks. The colours and magnitudes of
bulges and disks have been corrected for inclination. The top panel in Figure
\ref{fig:CMD_overlay} shows that ellipticals and pure disk galaxies
are well-separated in colour space (this is due in part to the
selection of ellipticals based on colour). The bottom panel shows that
classical bulge galaxies occupy the green valley, while pseudo-bulge
galaxies are predominately blue (as expected from the colour selection
of bulges). Figure \ref{fig:CMD_overlayB} shows the CMDs for bulges
and disks separately. Although the scatter in colour is large,
classical bulges lie on the red sequence. The large scatter,
especially toward redder colours, is probably due to the difficulty in
fitting the linearly scaled B+D model in the $g$ and $i$-bands. The scatter is
similar if we use $r-i$ for the colour. The lower panel in this figure
shows the disk components. Disks around any type of bulge have
significant overlap in colour. Disks around classical bulges are
redder than bulge-less and pseudo-bulge galaxies, in agreement with
observations that bulge and disk colour are correlated \citep{deJong1994,Peletier96,Wyse1997,Cameron09}.

\figCMDGreenValley

Using our sample of fitted galaxies, we can examine the morphological
properties of $z\approx 0$ galaxies as a function of position in the
colour magnitude distribution (CMD). Figure \ref{fig:CMD_greenvalley}
shows the 
colour magnitude diagram for our sample divided into absolute magnitude
bins. Following \citet{Mendez11,Baldry04,Strateva01}, we fit each
magnitude bin with a double Gaussian, and then fit the red sequence by
fitting a line to the red peaks of the double Gaussians. We define the
center of the green valley as a line parallel to the red sequence that
goes through the minimum of the CMD between $-20<M_r\leq -18.5$. The
equation for this line is $(g-r) = -0.025(M_r+20) + 0.611$. The
width of the green valley is taken to be $0.1$ in $g-r$, following
\citet{Mendez11}. Using this definition of the green valley, and
restricting our sample to $69,666$ galaxies with $-23<M_r<-17$, we
find $56\%$($38,673$) of our sample is blue, $17\%$(11600) green, and
$28\%$($19,393$) is red. Galaxies with S\'ersic fits make up
$44\%$($17,156$), $32\%$($3,734$), and $20\%$($3,971$)  of
the blue, green, and red galaxies, respectively. Ignoring these
galaxies, the median $B/T$ for blue, green, and red galaxies is
$0.00$, $0.27$, and $1.0$, respectively.

Unsurprisingly, $87\%$($12,738$) of the exponential disk
galaxies fall in the blue cloud, and $94\%$($9,086$) of the de
Vaucouleurs profile-fitted galaxies are red (this is expected due to
our selection criteria for ellipticals, see Figure
\ref{fig:CMD_overlay}). The bulge+disk galaxies span a wider colour
distribution (Figure \ref{fig:CMD_overlay}). The fractions of $n_b=4$ B+D
galaxies which are blue, green, and red are $34\%$($4,805$), $31\%$($4,337$), and
$35\%$(4869). For the $n_b=1$ galaxies the same fractions are
$59\%$(3852), $24\%$(1562), and $17\%$(1087). Taken together, the B+D
modeled galaxies dominate the 
green valley (see Figure \ref{fig:CMD_overlay}). If we examine only
galaxies with good B+D fits, we can separate them by colour into blue,
green, and red and examine their properties. This is shown in Figure
\ref{fig:bgr_BD_fits}\figbgrBDFits. Here, we have included all B+D modeled
galaxies, with their bulge profile selected as in
\S\ref{ssec:n1n4bulge}. However, the general trends remain the same
if we fit all the bulges with either an $n_b=1$ or $n_b=4$ S\'ersic
profile. The rightmost column in Figure \ref{fig:bgr_BD_fits} shows
the distributions of (inclination-corrected) disk colours for the
blue, green and red galaxies. Of the properties plotted, the disk
colour changes the most as galaxies move from blue to red. The
inclination-corrected B/T is shown in the top row. The blue galaxies
have smaller bulges with a median B/T of $0.15$, while the green and red
galaxies have larger bulges, with median values of $0.33$ and
$0.42$. The median disk colours of blue, green, and red galaxies are
$0.43$, $0.58$, and $0.67$, respectively. The fact that a substantial
number of galaxies on the red sequence remain disk dominated is in
agreement with recent other observations \citep{Bundy2006, Bundy2010}. It
suggests that galaxies transitioning to the red sequence may do so
through a combination of bulge growth (mergers) and disk fading. 

\section{Summary}
\label{sec:summary}
We present a set of 2-dimensional bulge+disk models for $71,825$
nearby ($0.003<z<0.05$) SDSS galaxies. We have fit $5$ different
models to each of the galaxies: an exponential profile, a de
Vaucouleurs profile, a S\'ersic profile, a de Vaucouleurs bulge plus
exponential disk, and and exponential bulge plus exponential disk. Our
models keep the 
bulge S\'ersic index fixed in order to limit the number of free 
parameters. By fitting classical bulges and ellipticals with de
Vaucouleurs profiles, we can more easily facilitate comparisons
between their properties. In order to obtain colours, we linearly
scale each component of the $r$-band model in $g$ and $i$. We have
chosen to use $r$ band SDSS images for the fitting, but have shown
that these results are consistent to $\sim10\%$ with those found in
$g$ and $i$ band. Our single component fits are in good agreement with 
the single component fits from SDSS and \citet{Blanton2005}. We
compare our 2-component fits to G09 and find our models produce
similar disk parameters, but that our bulges are typically larger
and brighter due to the inclusion of a free S\'ersic index and a bar in the G09
models. Nonetheless, our values for $B/T$ and their values for B$+$Bar$/$T
are comparable. 

We examine the systematic and statistical errors in our fits by
examining the properties of fits to mock image and fits to a set of 39 bright
galaxies from our sample. For both sets of images, we have
independently varied the resolution and the S/N. We first examined
single component S\'ersic fits. For galaxies with high
S\'ersic indices, the value of \Reff{} decreases(increases) with
decreasing resolution(S/N). The opposite occurs for galaxies with low
S\'ersic indices. Furthermore, the measured S\'ersic index is also
sensitive to the resolution and S/N; it decreases (especially for
high-$n$ galaxies) as galaxies are moved to higher redshift. For
single-component fits, the fractional error in \Reff{} is approximately
$1.5\times$ as large as the error in the flux for images with
$\Reff/\mathrm{PSF\ FWHM} \ga 0.5$. The relative fractional error in
\Reff{} grows rapidly for smaller galaxies, suggesting that $0.5$ FWHM 
is a reasonable lower limit on \Reff{} for a reliable fit. For
2-component fits, there are no clear trends in the fit parameters with
decreasing S/N and resolution. Generally, the disk parameters are more
robust than 
the bulge parameters, and the measured $B/T$ increases with redshift
for $n_b=1$ B+D models. However, we do not find that these trends are
robust or significant enough to justify adjusting the fits of high-$z$
galaxies to make them comparable to fits of low-$z$ galaxies.

Since each galaxy is fit with $5$ different models, we create an
algorithm for selecting the preferred model for a given
galaxy. Although this algorithm takes into account the $\chi^2$ values
for each model, it relies mainly on astrophysically motivated
constraints. A galaxy will be fit with a B+D model, if the following
are true: the two component fit is a statistically significant
improvement over any single component fit; the bulge flux and disk
fluxes are nonzero in the linearly-scaled models in $g$ and $i$-bands;
the bulge scale length is smaller than the disk scale length; and the
bulge flux 
dominates in the center of the galaxy. These requirements are
satisfied for $29\%$ of our sample. The remainder of the galaxies are
fit with a single component exponential ($23\%$), de Vaucouleurs
($13\%$), or S\'ersic profile ($36\%$). The last category includes the
(intrinsically) faintest galaxies in the sample as well as irregular
galaxies, strongly barred galaxies and galaxies with otherwise poor
model fits. The fact that the fraction of S\'ersic-modeled galaxies as
a function of absolute magnitude is independent of the apparent
magnitude limit of the sample, demonstrates the robustness of our fitter
and our categorization algorithm; the fits and categorizations are not
strongly magnitude-limited in our sample. 

In order to distinguish between classical bulges ($n_b=4$) and
pseudo-bulges ($n_b=1$), we have relied on the colour and flatness of
the bulge. We select classical bulges to lie on the red sequence
($g-r > 0.6$). We have not used $\chi^2$ values for the two
different B+D models because they are only significantly different for
$\sim 5\%$ of all the galaxies. This leaves a sample of $14042$
classical bulges and $6684$ pseudo-bulges. This division of bulges by
colour is a division based on the current star formation occurring in
those bulges. We also examine the Kormendy relation for bulges and
ellipticals and show that the majority of classical bulges lie along
to the elliptical galaxy Kormendy relation, while the pseudo-bulges do
not. This supports the use of colour as a division between the two
dynamically different bulges.

Given our sample of B+D galaxies, we examine their properties as a
function of total galaxy colour. We find that classical bulge galaxies
are evenly distributed among red, green, and blue while pseudo-bulge
galaxies are predominately blue, but both types are a significant
fraction of green valley galaxies ($70\%$ for galaxies brighter than
$16$ in $r$). Furthermore, the difference between red and green B+D
galaxies is predominately due to the disk becoming redder, not growth of
the central bulge.  

This work presents a large, homogeneous sample of B+D
galaxies to date. Nonetheless, it is evident from our S/N and resolution 
tests, that there are serious limitations to the quality of B+D fits
from ground-based data. In classifying our fits, we have tried to be 
conservative, only using a B+D where it makes statistical and
astrophysical sense. Despite these limitations, the size of our sample
allows us to create large subsamples of galaxies of different
morphological types. In future works, we plan to examine the mass
fractions of bulges (and ellipticals) and disks, at least for massive
galaxies where the contribution from S\'ersic-modeled galaxies is
small. We also plan to examine single morphological types in greater
detail. For example, we have found $\sim14,000$ galaxies with robust
classical bulges, which span a wide range of $B/T$, luminosity(mass),
stellar age, and environment. Because the sample is large, we can
examine correlations between environment and classical bulge
properties and attempt to answer questions about bulge (and disk)
formation, in a statistically meaningful way. 

\section*{Acknowledgments}

We thank Jenny Greene, and the referee, Luc Simard, for their valuable
comments on this work. This work is supported by the NSF grant AST0908368.

Funding for SDSS-III has been provided by the Alfred P. Sloan
Foundation, the Participating Institutions, the National Science
Foundation, and the U.S. Department of Energy Office of Science. The
SDSS-III web site is http://www.sdss3.org/. 

SDSS-III is managed by the Astrophysical Research Consortium for the
Participating Institutions of the SDSS-III Collaboration including the
University of Arizona, the Brazilian Participation Group, Brookhaven
National Laboratory, University of Cambridge, University of Florida,
the French Participation Group, the German Participation Group, the
Instituto de Astrofisica de Canarias, the Michigan State/Notre
Dame/JINA Participation Group, Johns Hopkins University, Lawrence
Berkeley National Laboratory, Max Planck Institute for Astrophysics,
New Mexico State University, New York University, Ohio State
University, Pennsylvania State University, University of Portsmouth,
Princeton University, the Spanish Participation Group, University of
Tokyo, University of Utah, Vanderbilt University, University of
Virginia, University of Washington, and Yale University.


\bibliographystyle{mn2e}
\bibliography{lackner_short}

\label{lastpage}
\end{document}